\documentclass[preprint2,twocolumn,times,tighten,twocolappendix]{aastex63}

\usepackage{color}
\usepackage{enumitem}
\usepackage{hyperref}
\usepackage{verbatim}
\usepackage{amsmath,amstext,mathrsfs}
\usepackage[all]{hypcap} 
\usepackage{afterpage}    
\usepackage{marginnote}
\usepackage{makecell}
\usepackage{subfigure}
\usepackage{multirow}
\usepackage{lineno}

\shorttitle{Turbulent Magnetic Field Amplification}
\shortauthors{Hu et al.}
\graphicspath{{./}{figures/}}

\begin{document}

\title{Turbulent Magnetic Field Amplification by the Interaction of Shock Wave and Inhomogeneous Medium
}

\email{sxu@ias.edu; yue.hu@wisc.edu}

\author[0000-0002-8455-0805]{Yue Hu}
\affiliation{Department of Physics, University of Wisconsin-Madison, Madison, WI 53706, USA}
\affiliation{Department of Astronomy, University of Wisconsin-Madison, Madison, WI 53706, USA}
\author[0000-0002-0458-7828]{Siyao Xu}
\affiliation{Institute for Advanced Study, 1 Einstein Drive, Princeton, NJ 08540, USA\footnote{Hubble Fellow}}
\author{James M.Stone}
\affiliation{Department of Astrophysical Sciences, Princeton University, Princeton, NJ 08544, USA}
\affiliation{Program in Applied and Computational Mathematics, Princeton University, Princeton, NJ 08544, USA}
\author{Alex Lazarian}
\affiliation{Department of Astronomy, University of Wisconsin-Madison, Madison, WI 53706, USA}
\affiliation{Centro de Investigación en Astronomía, Universidad Bernardo O’Higgins, Santiago, General Gana 1760, 8370993,Chile}

\begin{abstract}
Magnetic fields on the order of 100 $\mu$G observed in young supernova remnants cannot be amplified by shock compression alone. To investigate the amplification caused by turbulent dynamo, we perform three-dimensional MHD simulations of the interaction between shock wave and inhomogeneous density distribution with a shallow spectrum in the preshock medium. The postshock turbulence is mainly driven by the strongest preshock density contrast and follow the Kolmogorov scaling. The resulting turbulence amplifies the postshock magnetic field. The time evolution of the magnetic fields agrees with the prediction of the nonlinear turbulent dynamo theory in \cite{2016ApJ...833..215X}. When the initial weak magnetic field is perpendicular to the shock normal, the maximum amplification of the field's strength achieves a factor of $\approx200$, which is twice larger than that for a parallel shock. We find that the perpendicular shock exhibits a smaller turbulent Alfv\'en Mach number in the vicinity of the shock front than the parallel shock. However, the strongest magnetic field has a low volume filling factor and is limited by the turbulent energy due to the reconnection diffusion taking place in a turbulent and magnetized fluid. The magnetic field strength averaged along the $z$-axis is reduced by a factor $\gtrsim10$. We decompose the turbulent velocity and magnetic field into solenoidal and compressive modes. The solenoidal mode is dominant and evolves to follow the Kolmogorov scaling, even though the preshock density distribution has a shallow spectrum. When the preshock density distribution has a Kolmogorov spectrum, the turbulent velocity's compressive component increases. 
\end{abstract}


\section{Introduction}
Shock waves, such as supernova blast waves, the heliospheric termination shock, and fast shocks in the interstellar medium (ISM) are crucially and extensively involved in a variety of astrophysical processes \citep{1975ApJ...195..715M,1983ApJ...275..652S,1999SSRv...90..413R,2003MNRAS.339..133S,2005Sci...309.2017S,2007ApJ...663L..41G}. For instance, the high-density contrast and compressed filaments created by supersonic shocks in molecular clouds serve as the nurseries for new stars \citep{MK04,MO07,2018MNRAS.480.3916M,2021MNRAS.504.4354B,HLS21,LHL22}. 

The supernova shocks are responsible for accelerating charged particles to high energies via the most accepted mechanism of diffusive shock acceleration \citep{1969JGR....74.1301S,1977DoSSR.234.1306K,1978MNRAS.182..147B,1980ApJ...237..793B,2000IAUS..195..291A} or its combination with shock drift acceleration (see \citealt{2022ApJ...925...48X}). However, a weak magnetic field $\approx5$~$\mu$G as expected in the ISM (see \citealt{Crutcher04}) is not sufficient to confine and accelerate Galactic cosmic rays to Pev energies \citep{1983A&A...125..249L}. 

Magnetic field amplification is crucial for understanding the acceleration of cosmic rays  \citep{2019Ap&SS.364..185U,2020LRCA....6....1M,2022ApJ...925...48X}. Although the interstellar magnetic fields can be amplified by shock compression, it alone cannot explain the observed magnetic field strength on the order of 100~$\mu$G behind a young supernova blast wave \citep{2005ICRC....3..233V,2003ApJ...589..827B,2003A&A...412L..11B,2014ApJ...790...85R}. In particular, X-ray and radio observations of the supernova remnant Cassiopeia A revealed that the magnetic field can be amplified by a factor of $\approx100$ compared with those in ambient ISM \citep{1977MNRAS.179..573B,1995ApJ...441..300A,1995ApJ...441..307A,2000ApJ...537L.119H,2006NatPh...2..614S,2009ApJ...697..535P}.
Such a substantial enhancement of magnetic fields was earlier considered as a result of the Bell mechanism, which proposed that cosmic ray's streaming induces the instability of reacting current in background plasma and amplifies the magnetic field \citep{2004MNRAS.353..550B,2005MNRAS.358..181B,2013MNRAS.430.2873R,2018MNRAS.473.3394V}.
\begin{table*}
	\centering
	\label{tab.1}
	\begin{tabular}{| c | c | c | c | c | c | c | c | c | c |}
		\hline
		Model & $\alpha$ & $\langle|\delta\rho/\rho_0|\rangle$ & max\{$|\delta\rho/\rho_0|\}$ & $\pmb{B}_0$ direction & $M_{\rm A,0}^{\rm shock}$ & $M_{\rm s,0}^{\rm shock}$ & $\beta^{\rm pre}_0$ & max\{$\rho_{\rm max}/\rho_0$\} & max\{$|\pmb{B}_{\rm max}|/|\pmb{B}_0|$\}\\\hline\hline
		Parallel (Para.) & -0.50 & 0.13 & 0.74  & $x$-axis & 224 & 100 & 10 & 7.0 & 65\\
		Perpendicular (Perp.) & -0.50 & 0.13 & 0.74 & $y$-axis & 224 &100 & 10 & 7.0 & 210 \\
		Kolmogorov (Kol.) & -1.67 & 0.18 & 0.87 & $x$-axis & 224 &100 & 10 & 7.3 & 80 \\
		\hline
	\end{tabular}
	\caption{Setups of MHD simulations.
	"Para." and "Perp." correspond to parallel and perpendicular shocks with a preshock shallow density spectrum. "Kol" corresponds to a parallel shock with a preshock steep Kolmogorov density spectrum.
	$\langle|\delta\rho/\rho_0|\rangle$, max\{$|\delta\rho/\rho_0|\}$, $\pmb{B}_0$, and $\beta^{\rm pre}_0$ are initial preshock's mean density field contrast, maximum density contrast, and plasma compressibility, respectively. $M_{\rm A,0}^{\rm shock}$ and $M_{\rm s,0}^{\rm shock}$ are Alfv\'en and sonic Mach numbers of the shock. $|\pmb{B}_{\rm max}|$ and $\rho_{\rm max}$ represent the unaveraged maximum values of magnetic field and density among all snapshots, respectively. $\alpha$ is 1D power-law index of the preshock density spectrum. $\rho_0$ and $|\pmb{B}_0|$ are the initial uniform preshock density and magnetic field strength, respectively. Numerical values of initial magnetic field strength and mean preshock density are the same for all three models. 
	}
\end{table*}

Supernova shocks in the Galactic disk propagate in a highly inhomogeneous medium and can interact with dense circumstellar clouds \citep{1970ApJ...162..485V,2000ApJ...537L.119H}, where the density spectrum is non-Kolmogorov and characterized by small-scale large density contrasts \citep{2007ApJ...658..423K,2009SSRv..143..357L,2012A&ARv..20...55H,2017ApJ...846L..28X,2020ApJ...905..159X}. Naturally, the amplification of magnetic fields takes places via the turbulence driven by the shock and density inhomogeneous' interaction in both preshock and postshock regions \citep{2009ApJ...707.1541B,2012MNRAS.427.2308D,2014MNRAS.439.3490M,2017ApJ...850..126X}. Pre-shock turbulent dynamo was numerically studied in \cite{2016MNRAS.458.1645D}, where the cosmic ray precursor interacts with upstream density fluctuations. Here we focus on the post-shock turbulent dynamo, and the turbulence is driven by the interaction of the shock with upstream density fluctuations. When shock interacts with preexisting density inhomogeneities, the shock front ripples, causing considerable vorticity in the downstream fluid \citep{2005ApJ...634..390B,2007ApJ...663L..41G,2013ApJ...770...84F,2013ApJ...772L..20I}. The postshock magnetic field is amplified by the vorticity-driven turbulence via the (nonlinear) turbulent dynamo \citep{2016ApJ...833..215X}. Note in this work, turbulence particularly refers to turbulent velocity fluctuation.

The turbulent dynamo in the post-shock region has been predominantly investigated by a number of studies using 1D \citep{1989ApJS...70..497D} or 2D MHD simulations \citep{2007ApJ...663L..41G,2009ApJ...695..825I,2012ApJ...747...98G,2013ApJ...770...84F,2014MNRAS.439.3490M}. It was shown that turbulent dynamo could amplify the postshock magnetic field by a factor of $\sim 100$. The magnetic field's morphology in 2D simulations, however, is different from the one in more realistic 3D simulations, which was recently investigated by \cite{2013ApJ...772L..20I} and \cite{2016MNRAS.463.3989J}. The 3D studies adopted a Kolmogorov spectrum for the preexisting density fluctuations, which is observed in the warm diffuse ISM \citep{1995ApJ...443..209A,2010ApJ...710..853C}. For supernova shocks propagating through the highly inhomogeneous medium, a shallow density spectrum is widely observed \citep{1998A&A...336..697S,2000ApJ...543..227D,2006PhDT........19S,2009SSRv..143..357L,2012A&ARv..20...55H,2017ApJ...846L..28X,2018ApJ...856..136P,HLB20,2020ApJ...905..159X}. The effect of preshock shallow density spectrum on driving postshock turbulence and turbulent dynamo has not been carefully studied. Motivated by this observational fact, in this work, we investigate the post-shock magnetic field amplification by using 3D MHD simulations of a planar shock wave that propagates in density fluctuations with a shallow density spectrum. Specifically, the preshock density fluctuations are limited to only large scales (i.e., wavenumber $k<10$) for the convenience of studying turbulent cascade to smaller scales. Also, to focus on the effect of preshock density inhomogeneities on post-shock turbulence and turbulent dynamo, we do not consider preexisting preshock velocity fluctuations. In particular, we will numerically test the nonlinear turbulent dynamo theory developed by \cite{2016ApJ...833..215X} in the context of shocks. It takes into account the magnetic feedback on turbulence and can be generally applied to studying magnetic field amplification in diverse astrophysical processes. \cite{2016ApJ...833..215X}'s theory explains the inefficient dynamo growth seen in simulations \citep{2009ApJ...693.1449C} and has been previously tested with numerical simulations of magnetic field amplification during galaxy cluster formation and first star formation \citep{2021arXiv210807822S,2022MNRAS.511.5042S}.

This paper is organized as follows. In \S~\ref{sec:data}, we give the details of the simulation's setup and numerical method. In \S~\ref{sec:results}, we investigate the turbulent amplification of magnetic field by the interaction of shock wave and inhomogeneous preshock density distribution. We present our numerical results accordingly. In \S~\ref{sec:dis}, we discuss the implication and prospects of the magnetic field amplification by shock waves. We summarize our work in \S~\ref{sec:con}.
\begin{figure}[htp]
    \centering
	\includegraphics[width=1.0\linewidth]{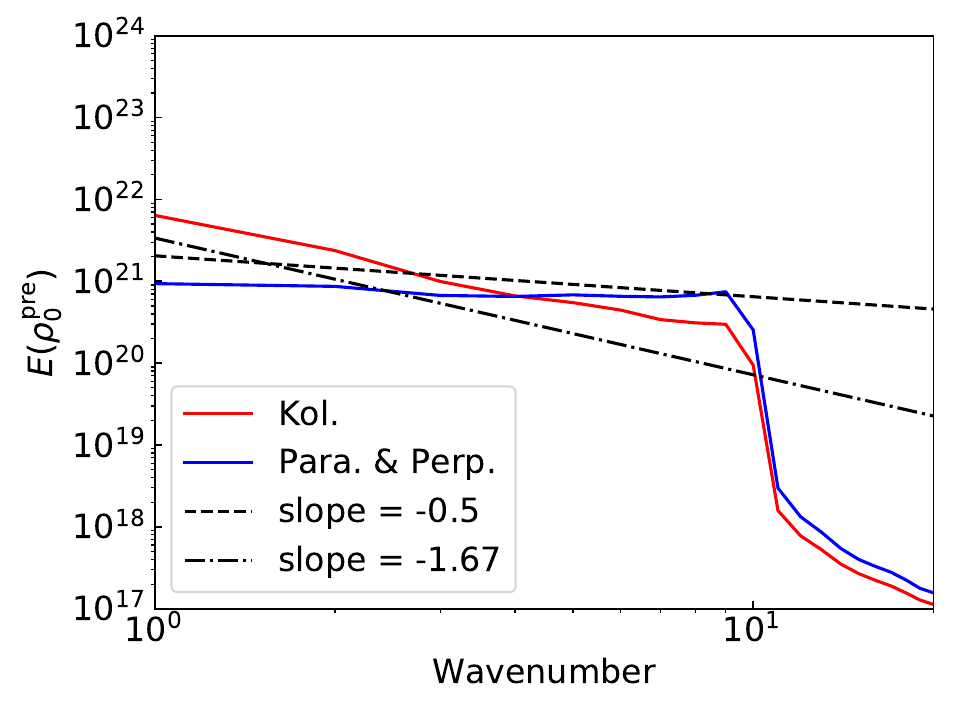}
    \caption{Spectra of initial preshock density fluctuations. Blue and red lines correspond to shallow and Kolmogorov density spectra, respectively.}
    \label{fig:pre.}
\end{figure}

\section{Numerical method}
\label{sec:data}

The MHD simulations used in this work are generated through Athena++ code \citep{Stone20}. We solve the ideal MHD equations of single fluid in the Eulerian frame: 
\begin{equation}
    \begin{aligned}
      &\partial\rho/\partial t +\nabla\cdot(\rho\pmb{v})=0,\\
      &\partial(\rho\pmb{v})/\partial t+\nabla\cdot[\rho\pmb{v}\pmb{v}+(P+\frac{B^2}{8\pi})\pmb{I}-\frac{\pmb{B}\pmb{B}}{4\pi}]=0,\\
      &\partial\pmb{B}/\partial t-\nabla\times(\pmb{v}\times\pmb{B})=0,\\
      &\nabla \cdot\pmb{B}=0,
    \end{aligned}
\end{equation}
where $\rho$ is gas density, $\pmb{v}$ is velocity, $P$ is gas pressure, and $\pmb{B}$ is magnetic field. We split the initial setup for postshock and preshock regimes to model a shock propagating in an inhomogeneous ISM and uniformly grid the box into $1024\times512\times512$ cells. On the upper and lower sides of the $x$-axis and $z$-axis, the boundary conditions are periodic, whereas gas can freely flow out from the sides of the $y$-axis. Several critical parameters are presented in Tab.~{\color{blue} 1} to characterize the scale-free simulation. The following sections detail our numerical settings. The superscripts $''$pre$''$ and $''$post$''$ are used to distinguish variables in preshock and postshock media and the subscript $''0''$ means initial value.

\subsection{Generation of density fluctuations in preshock medium}
For the preshock medium, we consider the density field in the form of $\rho^{\rm pre}_0=\rho_0+\delta\rho$, where $\rho_0$ is the uniform preshock density field, and $\delta\rho$ stands for zero-mean fluctuations. To create the fluctuation, we create a random density field in Fourier space with a power spectrum $\propto k^{-(\alpha+2)/2}$ and then transform it back into the real space. Explicitly, the density field in Fourier space is created as a realization of a Gaussian random field amplitude $A(k_n)$ and a uniform random field phase $\phi_n$. $\delta\rho$ is given by:
\begin{equation}
    \delta\rho = \sum_{n=1}^N A(k_n)e^{i(k_x x+k_y y + k_z z+\phi_n)},
\end{equation}
where $N$ is the total number of the modes, $k_n=2\pi/l=\sqrt{k_x^2+k_y^2+k_z^2}$ is the wavenumber ($k_x$, $k_y$, and $k_z$ are the $x$, $y$, $z$ components, respectively) and $l$ is the length scale in real space. We consider a shallow density spectrum with the 1D spectral index smaller than unity \citep{LP04,LP06}. As the shallow density spectra observed in the ISM have non-universal spectral indices \citep{2009SSRv..143..357L,2012A&ARv..20...55H}, we adopt $E_k(\rho^{\rm pre}_0)\propto k^{-\alpha} = k^{-1/2}$ as a representative example. The density spectrum is truncated at $k=10$ so that density fluctuations existing over $k<10$ (corresponding to length scale larger than $1/10L_x$; see Fig.~\ref{fig:pre.}). Consequently, the wavenumber's $x,y,z$ components $k_x, k_y, k_z$ range from 0 to 9. Such choice results in total $\approx600$ modes for $k<10$. As a shallow density spectrum has the correlation length of density fluctuations given by the inner scale, the largest density contrast appears around $k\sim 10$. We expect that the turbulence generated by the interaction of the shock with density fluctuations cascades down to smaller scales at $k>10$. As a comparison, we also consider a steep Kolmogorov density spectrum $E_k(\rho^{\rm pre}_0)\propto k^{-5/3}$ over $k<10$. The density fluctuations with a steep density spectrum, for which the 1D spectral index is larger than unity, has the correlation length given by the outer scale. We make its largest density contrast comparable to that of a shallow density spectrum. Particularly, if we assume that a supernova remnant (SNR) has a radius $\approx1$~pc \citep{2008Sci...320.1195K}, the injection scale of the background interstellar turbulence is $\approx100$~pc \citep{1995ApJ...443..209A,2010ApJ...710..853C}, and density fluctuation ($\delta \rho/\rho_0=1$ at the turbulence injection scale, \citealt{2007ApJ...665L..35D}) follows the Kolmogorov scaling, the density contrast at $\approx1$~pc is roughly $(1~{\rm pc} / 100~{\rm pc})^{1/3}\rho_0 \approx 0.2 \rho_0$. In the simulations, we have the mean ratio of density fluctuation and mean density $\approx0.15$ averaged over the entire preshock volume. As here our purpose is to investigate the effect of density spectral shape, we keep the largest density contrast similar for all cases. The effect of density contrast’s amplitude on the generated postshock turbulence was studied by \citet{2013ApJ...772L..20I}.


In this case of a parallel shock, the initial magnetic field $\pmb{B}^{\rm pre}_0=\pmb{B}^{\rm post}_0=\pmb{B}_0$ re-scaled with a factor of $1/\sqrt{4\pi}$ is parallel to the and shock normal, i.e., the $x$-axis. For a perpendicular shock, the setup of magnetic field strength is the same, but the direction is perpendicular to the shock normal. $|\pmb{B}_0|$ is determined by the plasma compressibility $\beta^{\rm pre}_0=2(c_{\rm s,0}^{\rm pre}/v_{\rm A,0}^{\rm pre})^2$ in the preshock medium. Here $c_{\rm s,0}^{\rm pre}=1$ (in numerical unit) is the initial sound speed of the preshock medium and $v_{\rm A,0}^{\rm pre}=|\pmb{B}_0|/\sqrt{\rho_0}$ is the initial Alfv\'en speed. We investigate two cases where the initial upstream magnetic field is parallel and perpendicular to the shock normal, i.e., parallel shock ("Para." in Tab.~1) and perpendicular shock ("Perp." in Tab.~1). To focus on the effects of preshock density spectral shape and shock obliquity, we keep all other parameters the same. The initial gas pressure $P^{\rm pre}_0$ in the preshock medium is given by the equation of state $P^{\rm pre}_0 = (c_{\rm s,0}^{\rm pre}\sqrt{\rho_0/\gamma})^2$, where $\gamma=5/3$ is the adiabatic index. 

\subsection{Injection of a planar shock wave}
The initial conditions in postshock medium are given by
the Rankine-Hugoniot relations \citep{1870RSPT..160..277M,hugoniot1889memoir} for adiabatic shock: 
\begin{equation}
\begin{aligned}
  \rho^{\rm post}_0&=\langle\rho^{\rm pre}_0\rangle\frac{\gamma+1}{(\gamma-1)+\frac{2}{M_{\rm s}^2}},\\
  P^{\rm post}_0&=P^{\rm pre}_0\frac{2\gamma M_{\rm s}^2-(\gamma-1)}{\gamma+1},\\
  |\pmb{v}^{\rm post}_0|&=|\pmb{v}^{\rm pre}_0|+M_{\rm s,0}\sqrt{\frac{\gamma P^{\rm pre}_0}{\langle\rho^{\rm pre}_0\rangle}}\frac{2(1-\frac{1}{M_{\rm s}^2})}{\gamma+1},\\
\end{aligned}
\end{equation}
where $M_{\rm s,0}$ is the sonic Mach number of the shock and $\pmb{v}^{\rm pre}_0=0$ is the initial velocity of preshock medium. $\langle...\rangle$ denotes ensemble average. Initially, the preshock and postshock magnetic fields are uniform. However, the magnetic field in the case of perpendicular shock will be compressed. The jump condition of the magnetic field is ${\pmb B}^{\rm post}={\pmb B}^{\rm pre}\frac{\gamma+1}{(\gamma-1)+\frac{2}{M_{\rm s}^2}}$.

The preshock density inhomogeneities are initialized to the entire simulation cube first and then we set up the initial uniform postshock (density, velocity, and pressure) fields (from $x=0$ to $x=L_x/80$) considering the jump conditions and set the shock front plane at $L_x/80$, where $L_x$ is the length of the $x$-axis. The shock propagates along the $x$-axis and its speed is given by $|\pmb{v}_{\rm shock}|=c_{\rm s,0}^{\rm pre}M_{\rm s,0}$. The shock propagation time to the end of the $x$-axis is $t_{\rm shock}\approx L_x/|\pmb{v}_{\rm shock}|$.

\begin{figure*}[htp]
    \centering
	\includegraphics[width=1.0\linewidth]{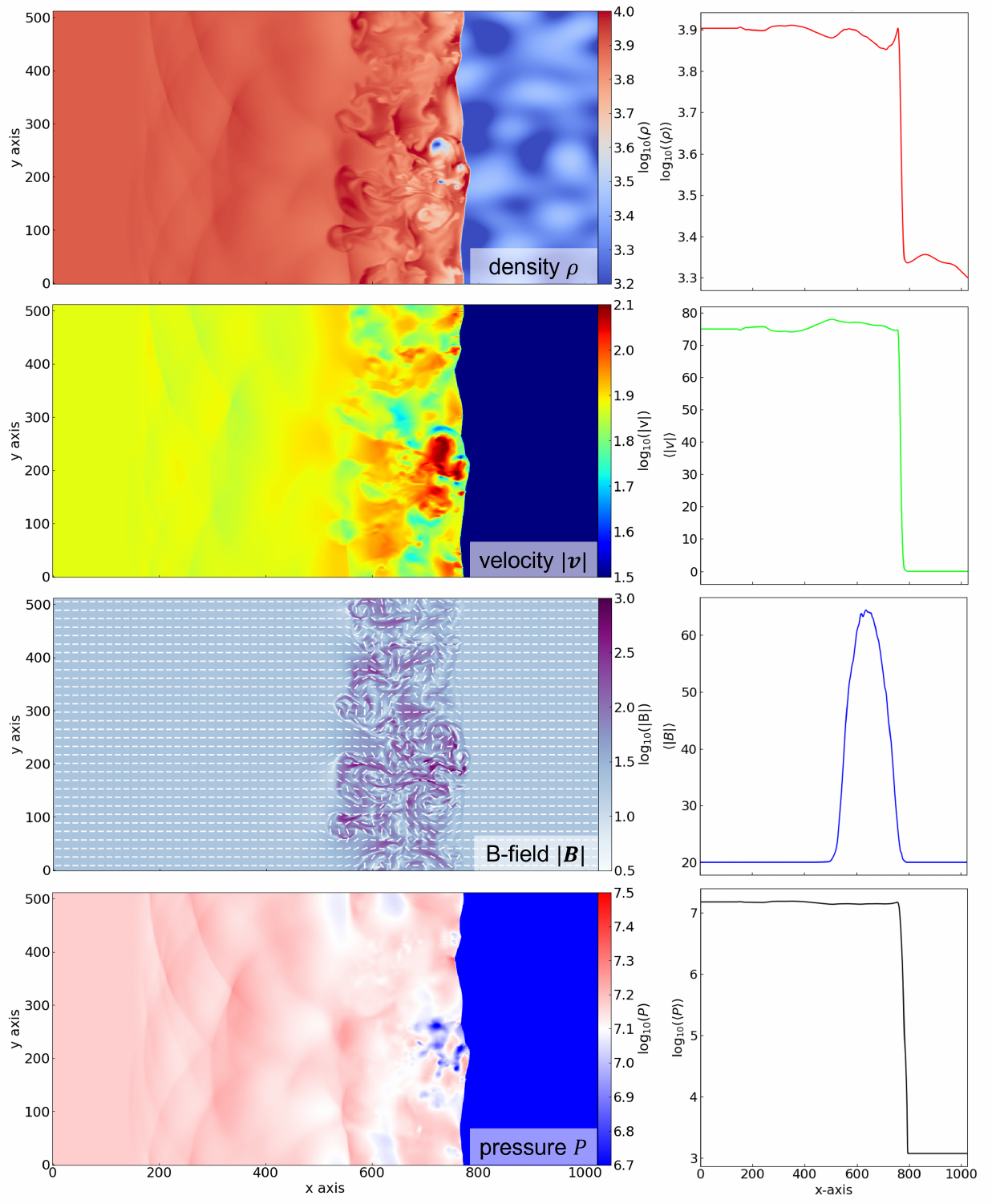}
    \caption{\textbf{Left:} the 2D cross sections of gas density, total velocity, magnetic field, and pressure at $z=256$. The snapshot of model "Para." at $t\approx0.75t_{\rm shock}$ is used here. The preshock mean magnetic field is parallel to the $x$-axis. Short white lines indicate the magnetic field orientations. \textbf{Right:} averaged (over $y$-axis and $z$-axis) profiles of the gas density, total velocity, magnetic field, and pressure as a function of $x$ using the same simulation. The physical variables are expressed in term of numerical units.}
    \label{fig:vis.}
\end{figure*}

\subsection{Helmholtz decomposition}
\label{ssec:Hel}
To examine the fraction of solenoidal component of turbulence, which accounts for the dynamo growth of magnetic fields, we adopt the Helmholtz theorem to decompose the turbulent velocity field into solenoidal component  $\pmb{v}_{\rm s}$ and compressive component $\pmb{v}_{\rm c}$:
\begin{equation}
\begin{aligned}
    \pmb{v}_{\rm tur}=\pmb{v}^{\rm post}-\pmb{v}^{\rm post}_0=\pmb{v}_{\rm s}+\pmb{v}_{\rm c}
\end{aligned}
\end{equation}
The solenoidal and compressive components satisfy divergence free ($\nabla \cdot \pmb{v}_{\rm s}=0$) and curl free ($\nabla \times \pmb{v}_{\rm c}=0$) conditions, respectively. Owing to the Helmholtz theorem, $\pmb{v}_{\rm c}$ stems from a scalar potential $\phi$, i.e., $\pmb{v}_{\rm c}=-\nabla\phi$, and $\pmb{v}_{\rm s}$ stems from a vector potential $\pmb{\Phi}$, i.e., $\pmb{v}_{\rm s}=\nabla\times\pmb{\Phi}$.

The two potentials can be calculated from the Green's function for the Laplacian:
\begin{equation}
\label{eq.green}
    \begin{aligned}
        \phi(\pmb{r})&=\frac{1}{4\pi}\int\frac{\nabla'\cdot \pmb{v}_{\rm tur}(\pmb{r}')}{|\pmb{r}-\pmb{r}'|}d\pmb{r'}^3,\\
        \pmb{\Phi}(\pmb{r})&=\frac{1}{4\pi}\int\frac{\nabla'\times \pmb{v}_{\rm tur}(\pmb{r}')}{|\pmb{r}-\pmb{r}'|}d\pmb{r'}^3,\\
    \end{aligned}
\end{equation}
where $\pmb{r}$ is the position vector and $\nabla'$ is the $\nabla$ operator with respect to $\pmb{r}'$. 
Thus, the decomposition can be rewritten as:
\begin{equation}
        \pmb{v}_{\rm tur}=-\frac{1}{4\pi}\nabla\int\frac{\nabla'\cdot \pmb{v}_{\rm tur}(\pmb{r'})}{|\pmb{r}-\pmb{r'}|}d\pmb{r'}^3+\frac{1}{4\pi}\nabla\times\int\frac{\nabla'\times \pmb{v}_{\rm tur}(\pmb{r'})}{|\pmb{r}-\pmb{r'}|}d\pmb{r'}^3.
\end{equation}
Note that Eq.~\ref{eq.green} basically is a convolution with the Green's function ($\frac{1}{4\pi|\pmb{r}-\pmb{r}'|}$). It takes advantages to solve Eq.~\ref{eq.green} in Fourier space, which is adopted in this work. The Fourier components of the potential fields are then transformed back into the real space to obtain the two velocity components. We decompose the turbulent magnetic field in a similar way, i.e., $\pmb{B}_{\rm tur}=\pmb{B}^{\rm post}-\pmb{B}^{\rm post}_0=\pmb{B}_{\rm s}+\pmb{B}_{\rm c}$.

\begin{figure*}
    \centering
	\includegraphics[width=1.0\linewidth]{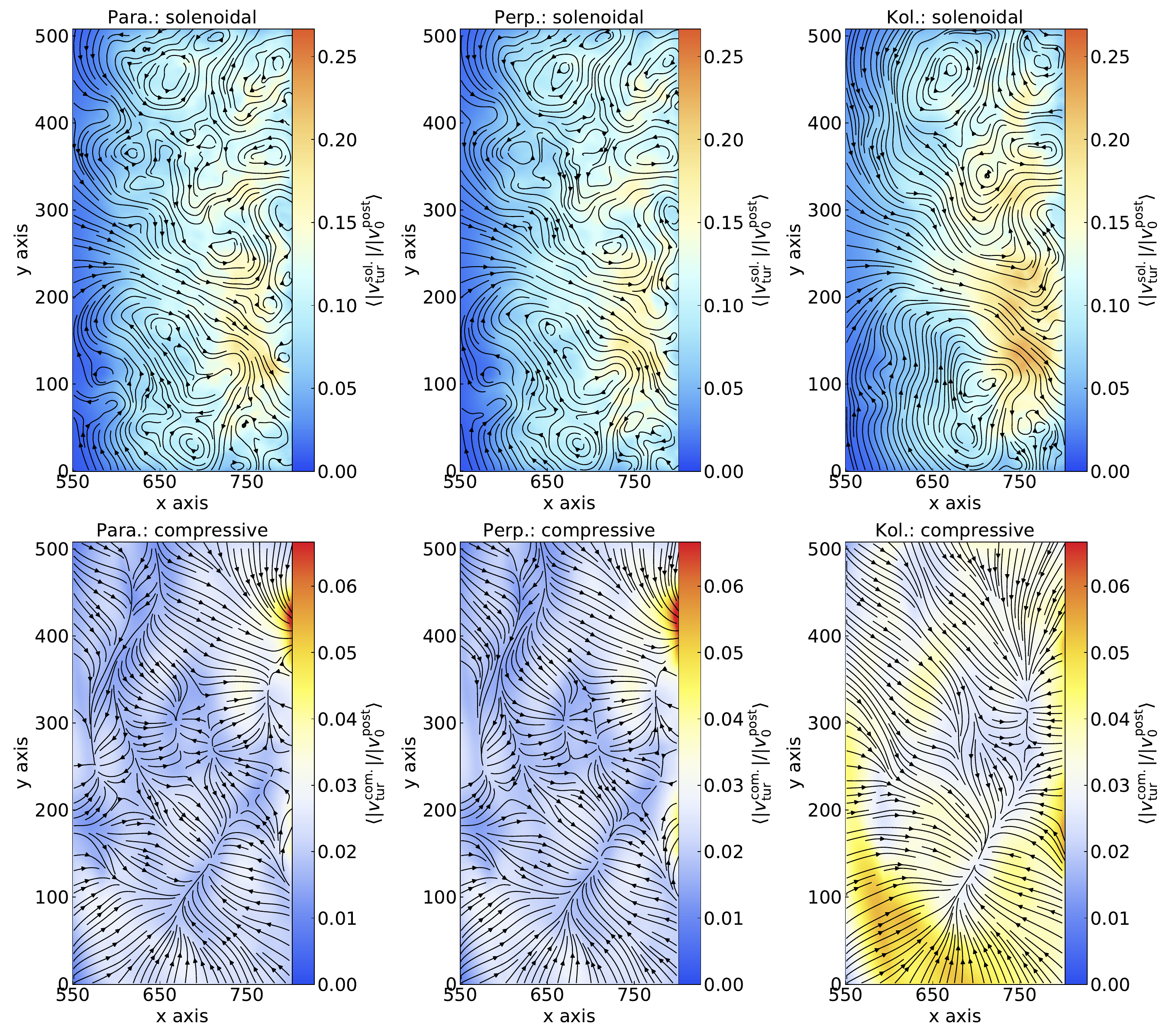}
    \caption{Distribution of projected decomposed turbulent velocity. \textbf{Top:} the solenoidal component averaged along the $z$-axis. \textbf{Bottom}: the compressive component averaged along the $z$-axis.The simulation cubes "Para." (left), "Kol." (middle), and "Perp." (right) at $t\approx0.75t_{\rm shock}$ are used here. The streamlines indicate velocity directions.}
    \label{fig:vsvc2d}
\end{figure*}

\begin{figure*}
    \centering
	\includegraphics[width=1.0\linewidth]{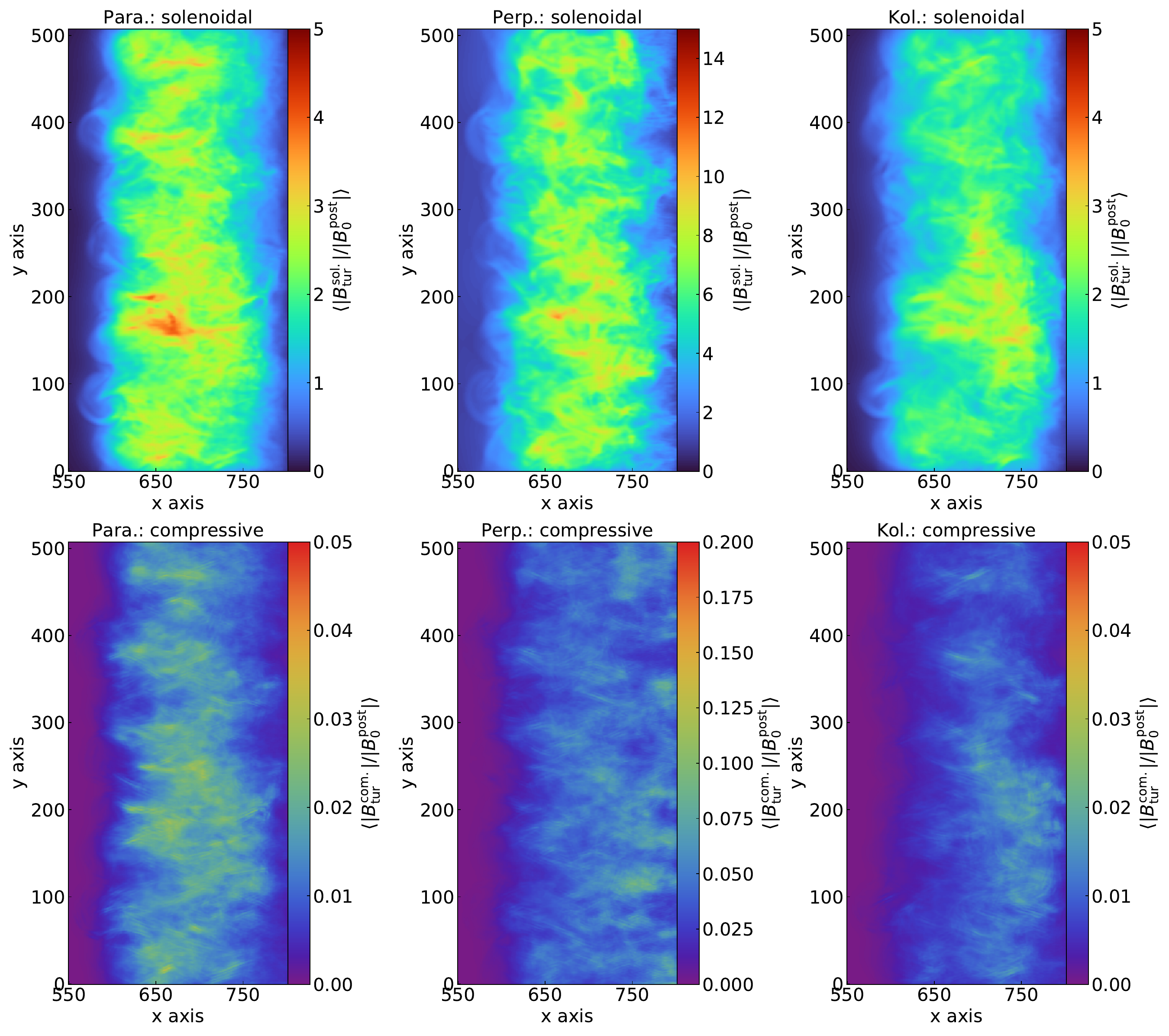}
    \caption{Distribution of projected decomposed turbulent magnetic fields. \textbf{Top:} the solenoidal component averaged along the $z$-axis. \textbf{Bottom}:  the compressive component averaged along the $z$-axis.The simulation cubes "Para." (left), "Kol." (middle), and "Perp." (right) at $t\approx0.75t_{\rm shock}$ are used here.}
    \label{fig:bsbc2d}
\end{figure*}

\section{Numerical results}
\label{sec:results}
\begin{figure*}[h!]
    \centering
	\includegraphics[width=1.0\linewidth]{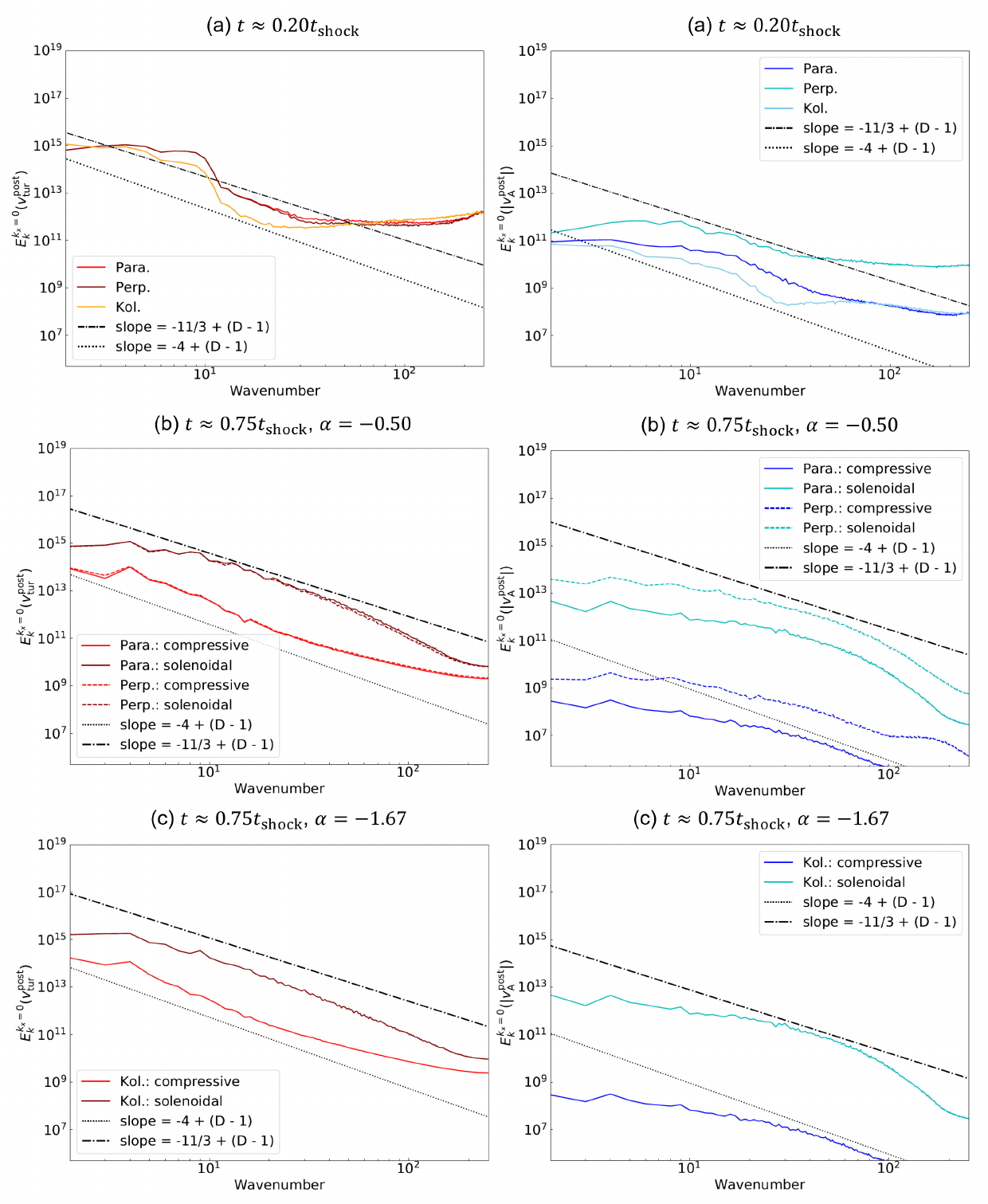}
    \caption{Panel (a): Energy spectrum of un-decomposed postshock turbulent velocity (left) and Alfv\'en velocity (i.e., magnetic field; right) at $t\approx0.20t_{\rm shock}$. Panel (b): Energy spectrum of decomposed postshock turbulent velocity (left) and magnetic field (right) at $t\approx0.75t_{\rm shock}$ for the cases of shallow preshock density spectrum. Panel (c): same as (b), but for the case of Kolmogorov density spectrum. Note we consider the energy spectrum only on the $k_x=0$ plane in Fourier space. Therefore, we have the slope $\approx -4 + (D-1)$ and $\approx -11/3 + (D-1)$ for compressive and solenoidal turbulence, respectively. Here $D=2$ stands for the case of 2D spectrum. $\alpha$ is the slope of the preshock density distribution's spectrum. }
    \label{fig:bvspectrum}
\end{figure*}

\begin{figure}[htp]
    \centering
	\includegraphics[width=1.0\linewidth]{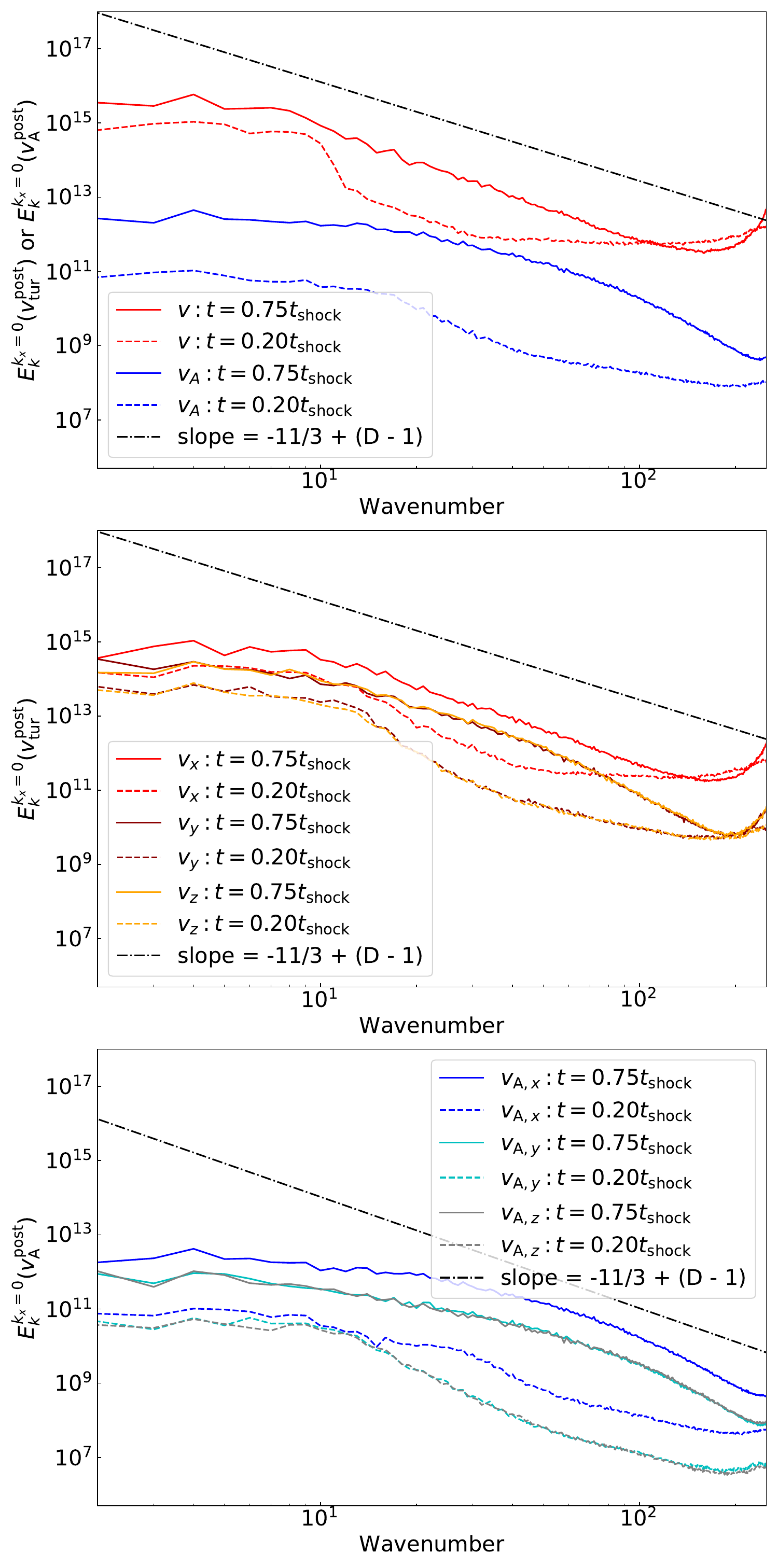}
    \caption{Energy spectrum of un-decomposed postshock turbulent velocity and Alfv\'en velocity using the "Para." model. Note we consider the spectrum only on the $k_x=0$ plane in Fourier space. Therefore, we have the slope $\approx -11/3 + (D-1)$ for solenoidal turbulence. Here $D=2$ stands for the case of 2D spectrum. }
    \label{fig:anisotropy_ratio}
\end{figure}

\begin{figure}[htp]
    \centering
	\includegraphics[width=1.0\linewidth]{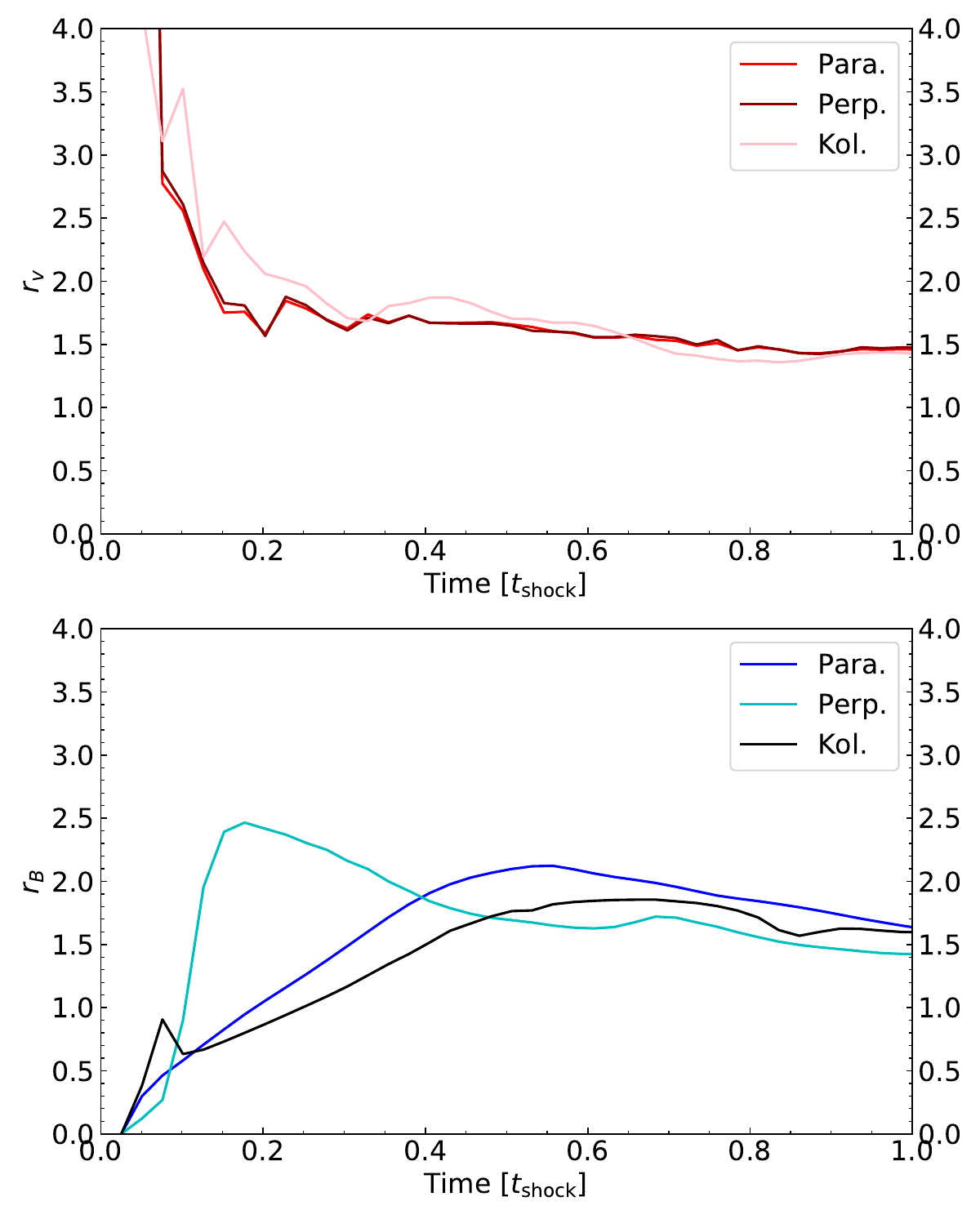}
    \caption{The anisotropy degree of velocity and magnetic field fluctuations. Here we define $r_v=2\delta v_x/(\delta v_y+\delta v_z)$ and $r_B=2\delta B_x/(\delta B_y+\delta B_z)$. The fluctuations are calculated over the postshock's turbulent shell.}
    \label{fig:anisotropy_ratio2}
\end{figure}

\begin{figure}[htp]
    \centering
	\includegraphics[width=1.03\linewidth]{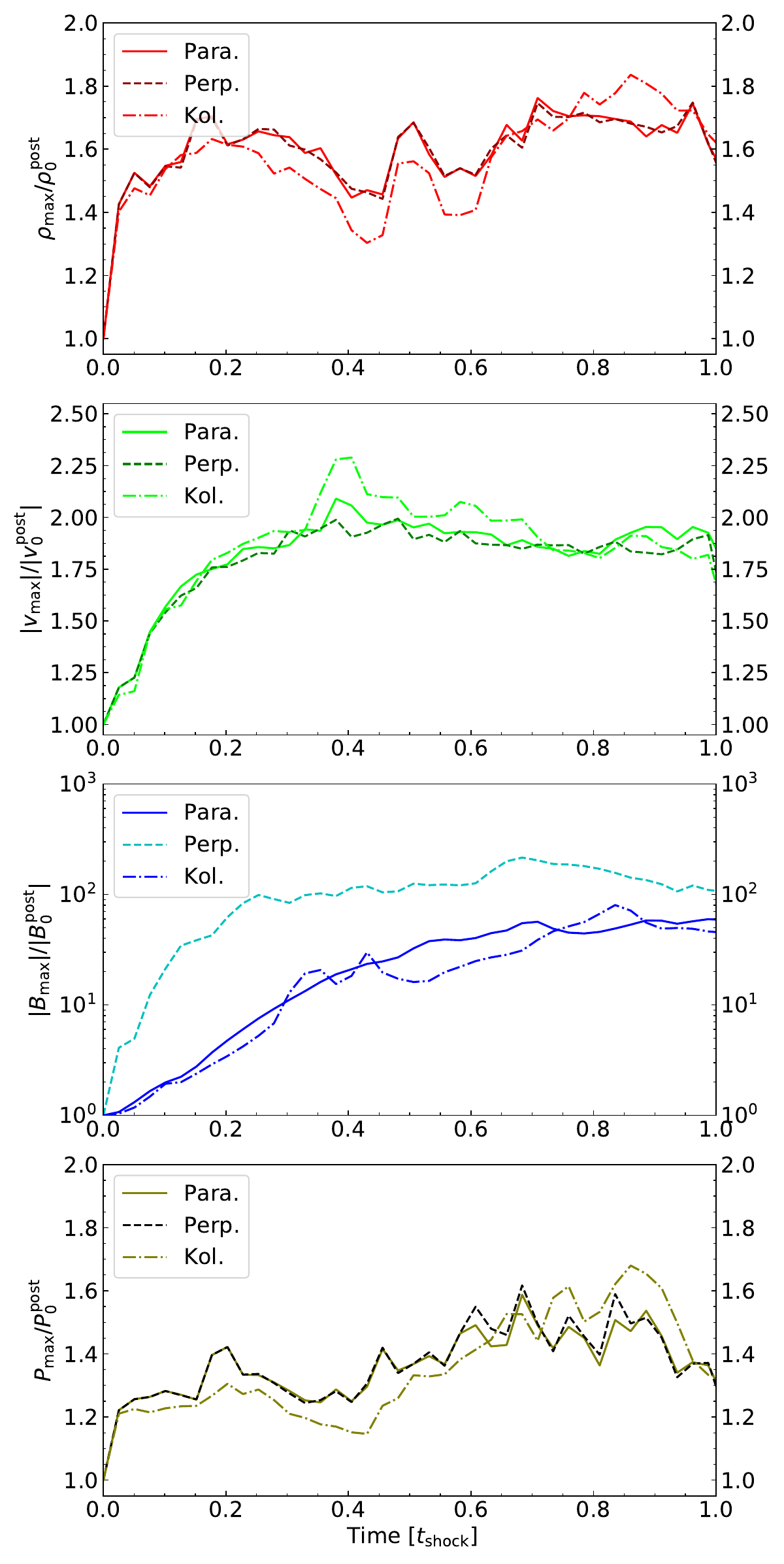}
    \caption{The ratio of the maximum density/(total) velocity /magnetic field/pressure and its initial value in the postshock region as a function of time.}
    \label{fig:max_time}
\end{figure}

\begin{figure}[htp]
    \centering
	\includegraphics[width=1.0\linewidth]{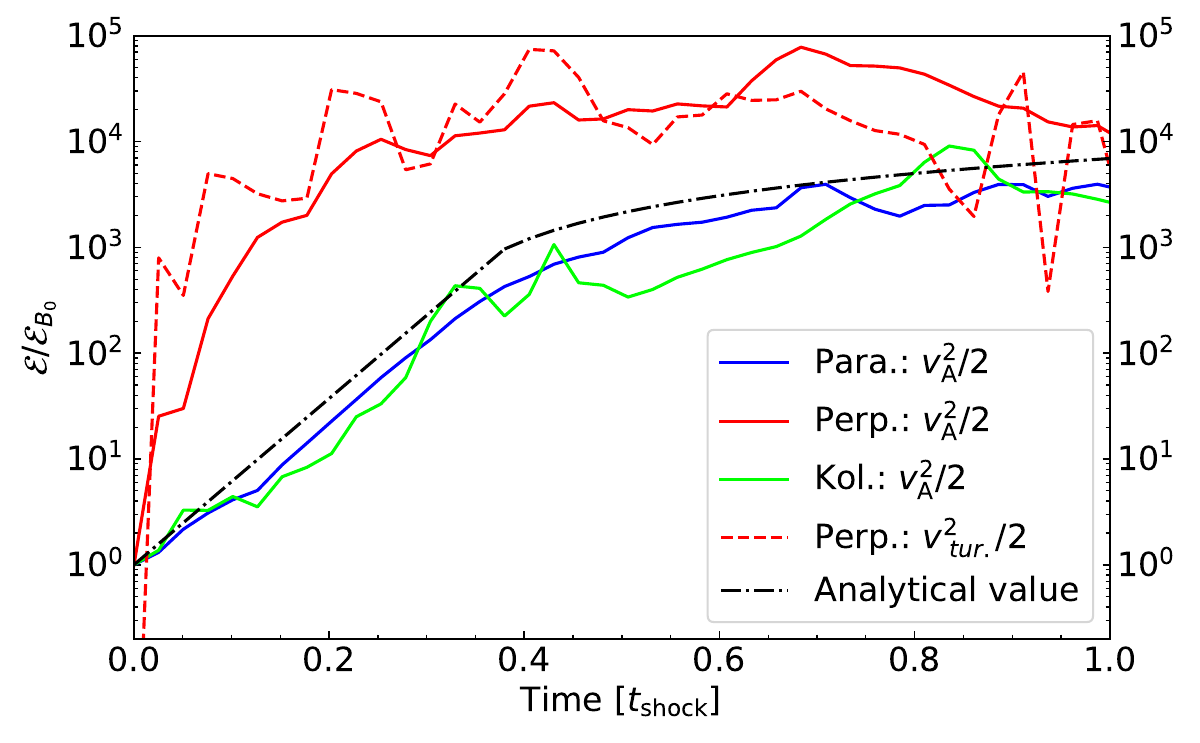}
    \caption{Magnetic field energy per unit mass $\mathcal{E}_B=v_{\rm A}^2/2$ (solid line) and turbulent kinetic energy per unit mass $\mathcal{E}_{\rm tur.}=v_{\rm tur}^2/2$ (dashed line) as a function of time. The calculation uses the maximum magnetic field strength and its corresponding density and turbulent velocity. The energies are normalized by the initial magnetic field energy $\mathcal{E}_{B_0}$. The dash-dotted line denotes the analytical prediction derived in \cite{2016ApJ...833..215X}.}
    \label{fig:energy}
\end{figure}

\begin{figure}[htp]
    \centering
	\includegraphics[width=1.03\linewidth]{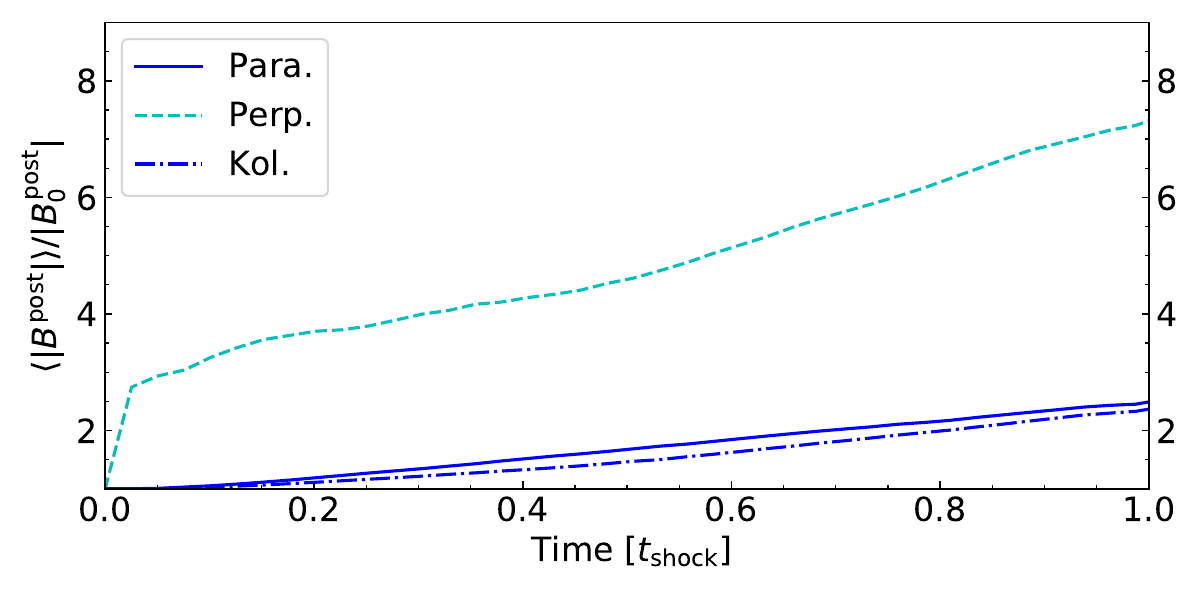}
    \caption{The ratio of averaged magnetic field and its initial value over the postshock's turbulent shell as a function of time.}
    \label{fig:mean_time}
\end{figure}

\begin{figure}[htp]
    \centering
	\includegraphics[width=1.0\linewidth]{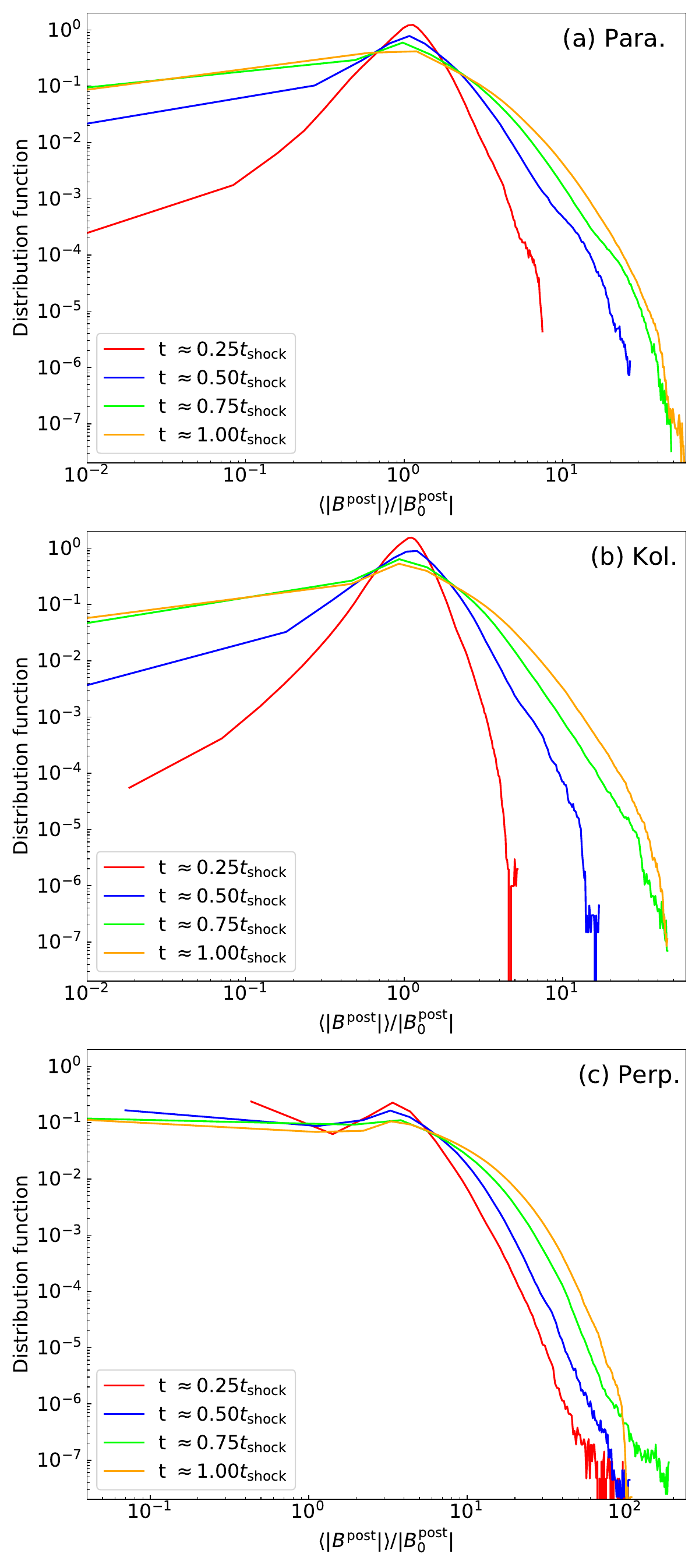}
    \caption{The histogram of the amplified magnetic field strength in the postshock's turbulent shell.}
    \label{fig:hist}
\end{figure}

\begin{figure*}[htp]
    \centering
	\includegraphics[width=0.99\linewidth]{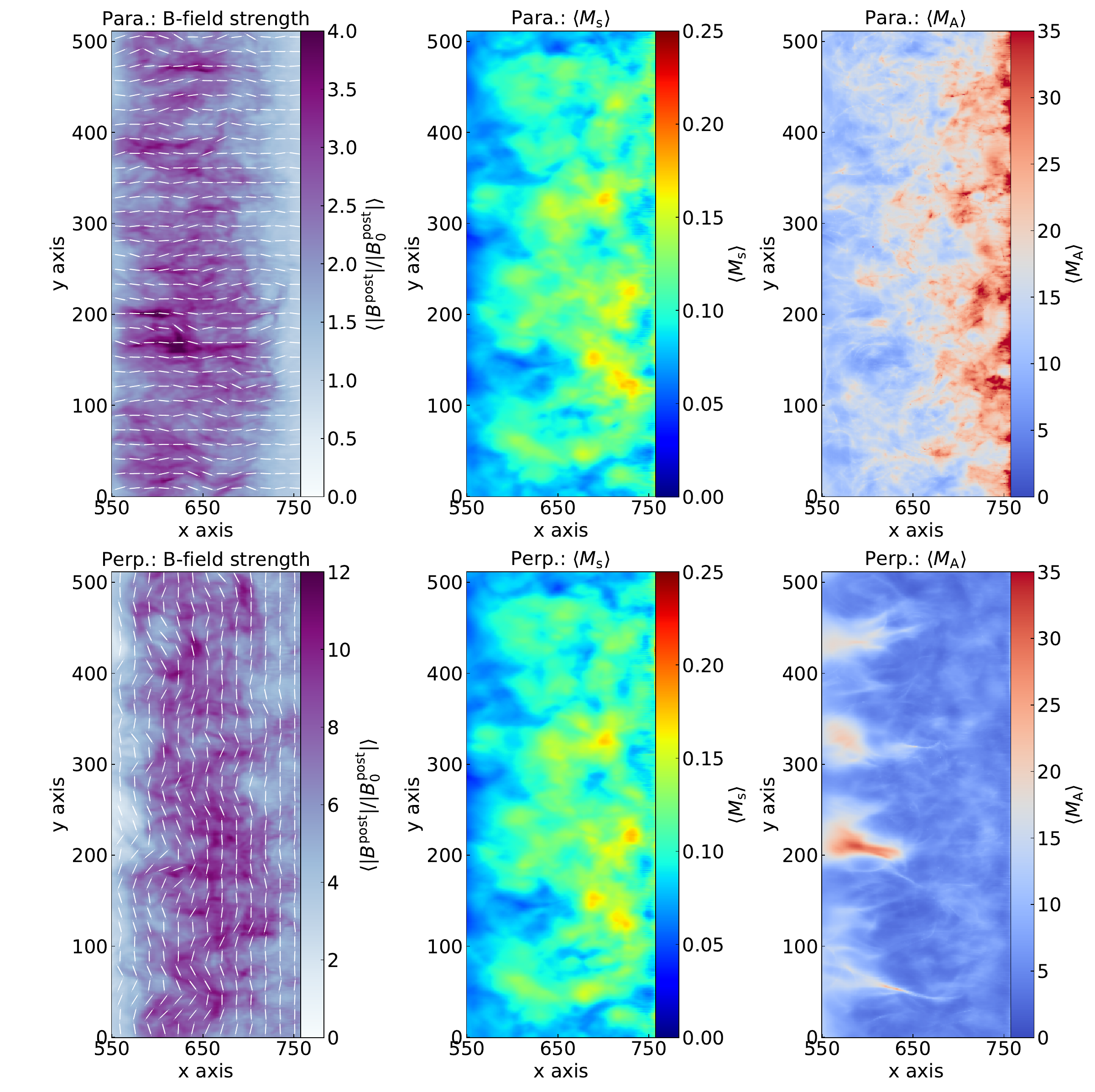}
    \caption{The distributions of averaged (along the $z$-axis) postshock magnetic field strength, $\langle M_{\rm s}\rangle$, and $\langle M_{\rm A}\rangle$ in the vicinity of shock front. $M_{\rm s}$ is the ratio of total velocity and sound speed, while $ M_{\rm A}$ is the ratio of turbulent velocity and Alfv\'en speed. The snapshot reaches $t\approx0.75t_{\rm shock}$. Short white lines indicate the magnetic field orientations.}
    \label{fig:mams}
\end{figure*}

\begin{figure}[htp]
    \centering
	\includegraphics[width=1.0\linewidth]{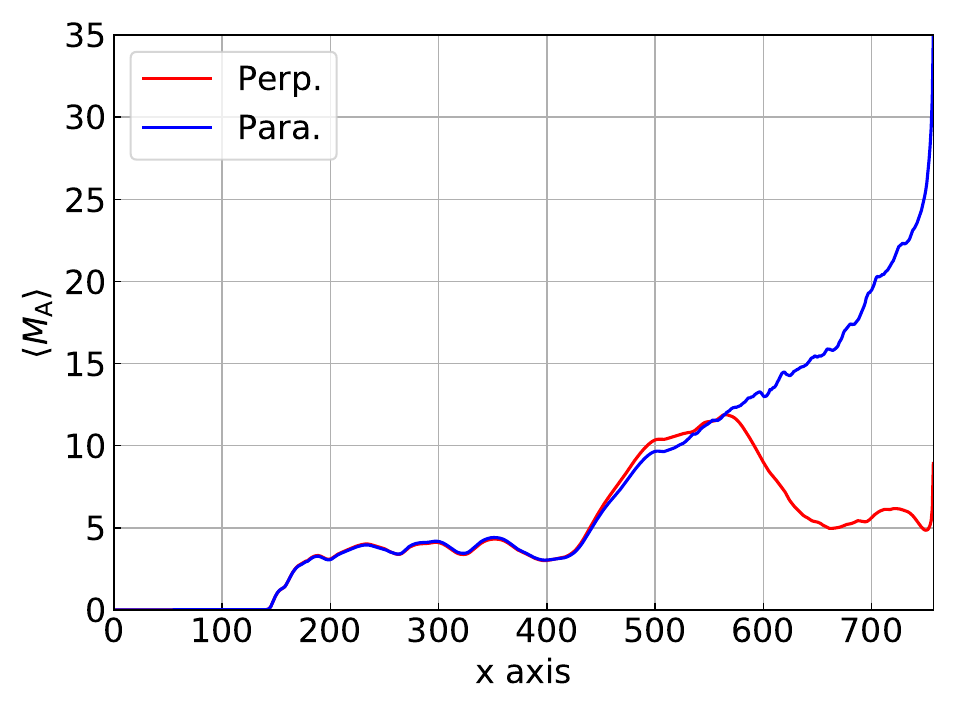}
    \caption{Averaged (over $y$-axis and $z$-axis) profiles of $\langle M_{\rm A}\rangle$ as a function of $x$ using the snapshot $t\approx0.75t_{\rm shock}$. Dashed line represents the left boundary of $\langle M_{\rm A}\rangle$'s maps shown in Fig.~\ref{fig:mams}.}
    \label{fig:ma}
\end{figure}

\subsection{Turbulence generated behind the shock front}
\label{ssec:turshe}
Fig.~\ref{fig:vis.} presents a visualization of a shock wave propagating in an inhomogeneous medium. It shows the structure of gas density, total velocity, magnetic field, and pressure at cross-section $z=256$. The simulation cube "Para." at $t\approx0.75t_{\rm shock}$ is used. Our results reveal that the four postshock quantities ($\rho$, $\pmb{v}$, $\pmb{B}$, and $P$) all develop fluctuations when the shock wave interacts with a fluctuating density field, and the shock front is corrugated. Their fluctuations are most significant in the shell with the thickness $\approx L_x/4$ behind the shock surface. The shell thickness is mainly determined by the largest turbulent eddy's turnover time $t_{\rm eddy}=L_{\rm inj}/v_{\rm inj}\approx 0.25t_{\rm shock}$, where $L_{\rm inj}\approx L_x/5$ is the effective injection scale\footnote{Turbulence here is driven at multiple scales from $k=1$ to $k=10$. We use $k=5$ as the effective injection scale.} of turbulence and $v_{\rm inj}\approx(v_{\rm max}-v_0^{\rm post})\approx 0.9v_0^{\rm post}\approx0.68v_{\rm shock}$ is the largest velocity fluctuation generated behind the shock (see Figs.~\ref{fig:bvspectrum} and \ref{fig:max_time}). Moreover, the structure of the amplified magnetic field appears filamentary. This suggests that the postshock magnetic field is enhanced by the turbulent stretching, i.e., turbulent dynamo. The reverse shock fronts are simply identified as the ripples with reversal motion with speed larger than sound speed in the postshock region. Their interactions are also observed farther downstream behind the shell. Their effect on magnetic fields is insignificant. Strong magnetic fields are only seen within the shell. We also present the magnetic field orientations. The initially ordered magnetic fields aligned along the $x$ axis become highly turbulent with random orientations in the shell.
 
In addition, we plot the profiles of averaged gas density, total velocity, magnetic field, and pressure as a function of $x$. The average is performed over the $y$ and $z$ directions perpendicular to the shock normal (i.e, the $x$ direction). As shown in Fig.~\ref{fig:vis.}, the shock wave compresses the preshock density and pressure by a factor of $\approx4$ and $\approx 10^4$, respectively, which is consistent with the strong shock limit ($M_{\rm s}\gg1$) of the jump conditions \citep{1870RSPT..160..277M,hugoniot1889memoir}. We find that when the postshock density fluctuation is taken into account, the postshock density can locally be $\approx7$ times higher than the preshock density (see Tab.~1).

Figs.~\ref{fig:vsvc2d} and \ref{fig:bsbc2d} present the maps of decomposed turbulent velocity and turbulent magnetic field at $t\approx0.75t_{\rm shock}$, respectively. The maps are averaged along the $z-$axis. The turbulent velocity is the similar for both parallel and perpendicular shock cases.  This shows that the generation of turbulence and its properties are insensitive to the shock obliquity given our numerical setups. Turbulence is induced by the Richtmyer-Meshkov instability (RMI; \citealt{2010RSPTA.368.1769N}). The shock front gets deformed when it interacts with a pre-shock density fluctuation. The basic mechanism for RMI is baroclinic vorticity generation caused by the misalignment of the pressure gradient and the local density gradient across the front \citep{2002AnRFM..34..445B}. As the shock front becomes more distorted, secondary instabilities, such as the Kelvin-Helmholtz instability \citep{1982JGR....87.7431M}, ultimately drive the postshock turbulence \citep{2013ApJ...770...84F,2016MNRAS.463.3989J}. The postshock turbulence is, therefore, regulated only by the pre-shock density fluctuations, consistent with \cite{2012ApJ...744...71I}. We find obviously the solenoidal component is dominant over the compressive component. The difference appears in the case of the Kolmogorov preshock density spectrum. We find a slightly larger compressive component compared to the case with a shallow density spectrum. As for the magnetic fields, the dominance of the solenoidal component is even more significant, with the solenoidal component's amplitude $\sim 100$ times higher than the compressive component. This is naturally expected as only the solenoidal turbulence is responsible for turbulent dynamo, which is much more efficient than the magnetic field amplification via turbulent compression.  Note that the magnetic field amplification is also correlated with the preshock density contrasts. A larger density contrast results in more significant postshock turbulence (i.e., velocity fluctuation) and consequently stronger amplification of magnetic fields \citep{2013ApJ...772L..20I,2013ApJ...770...84F}. 

\subsection{The 2D energy spectrum}
Fig.~\ref{fig:bvspectrum} shows the energy spectrum of total (i.e., including $x$, $y$, and $z$ components) turbulent velocity and Alfv\'en velocity, i.e., total magnetic field over the square root of density, at different time. To avoid the effect of non-periodic boundary conditions at the $x$-axis' left and right ends, we consider the spectrum only on the $k_x=0$ plane in Fourier space. The velocity spectrum exhibits a flat shape at $k\lesssim10$ and steepens toward larger $k$. The numerical dissipation effect becomes significant at $k\lesssim100$. As $k=10$ is the cutoff wavenumber of the preshock density spectrum, this suggests that the post-shock turbulence is driven at multiple scales at $k \lesssim 10$, and then cascades down to smaller scales at $k\gtrsim10$. Particularly, with the cutoff at $k=10$, we do not expect to see the driving effect on the turbulent spectral shape over $k>10$ (see Fig.~\ref{fig:bvspectrum}(a), left). The turbulence covers two orders of magnitude in length scales in our simulations. This range is similar to that of the magnetic energy spectrum in Tycho’s SNR measured by \cite{2018MNRAS.480.2200S}. The model "Kol." has a lower amplitude of turbulent spectrum. This suggests that with a similar density contrast, a preshock shallow density spectrum is more efficient in generating the postshock turbulence than a steep density spectrum. The slope of turbulent energy spectrum at $k\gtrsim10$ is between $\approx-8/3$ and $\approx-3$ and is independent of the preshock density spectral shape. The turbulent eddies with the largest eddy turnover rate dominate the dynamo amplification. Given that turbulence is mainly driven at $k< 10$ and cascade to smaller scales, the growth of postshock magnetic energy starts near from the smallest eddy at dissipation scale. Due to the additional magnetic field amplification via the shock compression, the model "Perp." leads to a larger amplitude of the magnetic energy spectrum. We find that as the shock surface is corrugated after interacting with upstream multiscale density fluctuations, the shock compression of magnetic fields happens at multiple scales over $k<10$.

In Fig.~\ref{fig:bvspectrum}, we further present the energy spectra of decomposed turbulent velocity and Alfv\'en speed at a later time $t = 0.75 t_\text{shock}$, where the dynamo is in the nonlinear stage (see Section \ref{subsec:ratio} and Fig.~\ref{fig:energy}). As we discussed earlier, the turbulence properties are insensitive to the shock obliquity. Given the same preshock shallow density spectrum, both parallel and perpendicular shocks have a very similar turbulent energy spectrum, and the solenoidal component dominates over the compressive component. For the fully developed turbulence (i.e., velocity fluctuation gets saturated, see Fig.~\ref{fig:max_time})., the spectral slope is independent of the preshock density spectral shape, i.e., the slope is the same for both cases of preshock density being shallow and Kolmogorov. The spectral slope for compressive velocity is close to $\approx-3$, while the slope for solenoidal velocity follows Kolmogorov scaling $\approx-8/3$. The magnetic energy spectrum is similar to the one of turbulence. The compressive and solenoidal cases have slopes $\approx-3$ and $\approx-8/3$, respectively. However, the magnetic energy spectrum at large scales (i.e., $k\lesssim10$) is flat and the spectral amplitude is lower than that of the turbulent energy spectrum. It shows that the full saturation of turbulent dynamo, i.e., the equipartition between the turbulent energy and magnetic energy at large scales, is not reached.

Fig.~\ref{fig:anisotropy_ratio} presents the energy spectra of total turbulence velocity and Alfv\'en speed,  as well as their $x$, $y$, $z$ components. Compared with the earlier exponential stage $t\approx0.2t_{\rm shock}$, the total (i.e., including all $x$, $y$, and $z$ components) magnetic energy, i.e., the Alfv\'en speed spectrum, and total turbulent energy increases further at the later non-linear stage $t\approx0.75t_{\rm shock}$. The magnetic energy is always smaller than the turbulent energy. 

In addition, we calculated the anisotropy degree of turbulence and magnetic fluctuations over the postshock's turbulent shell defined in Fig.~\ref{fig:mean_time}. The anisotropy degree is defined as the ratio of twice $x$-component's fluctuation to the sum of $y$ and $z$-components' fluctuations. Explicitly, for turbulent velocity, the anisotropy degree is: $r_v=2\delta v_x/(\delta v_y+\delta v_z)$, where $\delta v_x=\sqrt{\delta (v_{{\rm tur},x}^{\rm post})^2}$, $\delta v_y=\sqrt{\delta (v_{{\rm tur},y}^{\rm post})^2}$, and $\delta v_z=\sqrt{\delta (v_{{\rm tur},z}^{\rm post})^2}$. The anisotropy degree for magnetic field is defined in a similar way: $r_B=2\delta B_x/(\delta B_y+\delta B_z)$, where $\delta B_x=\sqrt{\delta (b_{{\rm tur},x})^2}$, $\delta B_y=\sqrt{\delta (B_{{\rm tur},y})^2}$, and $\delta B_z=\sqrt{\delta (B_{{\rm tur},z})^2}$. The results are presented in Fig.~\ref{fig:anisotropy_ratio2}. $r_v$ is greater than 1.5 when $t<0.2t_{\rm shock}$ suggesting the turbulence is most significant along the $x$ direction. After $t>0.2t_{\rm shock}$, the turbulence's different components gets sufficiently developed and a saturation value $\approx1.5$ is achieved. The trend of $r_B$, however, is opposite. $r_B$ starts from 0 and only after the turbulence's anisotropy gets saturated, the magnetic field's increment of all components gradually reach a stable state. In our case, $r_B$ gets saturated at $\approx1.75$ suggesting magnetic field's fluctuation along the $y$ and $z$ directions is smaller. Due to additional shock compression of the $y$ and $z$ components, the perpendicular shock case has a slightly smaller $r_B$ value, i.e., less anisotropic, than the case of a parallel shock. The results agree with the earlier study by \cite{2013ApJ...772L..20I}. They showed both the anisotropies of velocity and magnetic field fluctuations saturated at $\approx1.3$. As for the anisotropy ($r_B<1$) at initial stage, the magnetic field is perturbed and fluctuations appear in all $x$, $y$, and $z$ directions. The $y$ and $z$ components is further compressed by shock so that we observed $r_B<1$ when the compression effect dominates.

\subsection{Turbulent dynamo}
\label{subsec:ratio}
In this subsection, we quantify the evolution of density, total velocity, magnetic field, and pressure in the postshock region. In Fig.~\ref{fig:max_time}, we plot the ratio of maximum density/(total) velocity /magnetic field/pressure and its initial value in the postshock region as a function of time for three models "Para.", "Perp.", "Kol.". The time evolution of density, total velocity, and pressure basically don't show clearly difference between cases of parallel and perpendicular shocks. There is some difference seen between cases with different upstream density spectra. We find that the postshock density and velocity can be amplified by a factor of $\approx1.6$ on average, i.e., the ratio of maximum fluctuation and mean value is 1.6 times larger than the mean value. The maximum postshock density is seven times larger than the preshock density. The maximum value of postshock velocity depends on the preshock density fluctuations' amplitude, which act as the driving source of turbulence (see \S~\ref{ssec:turshe}). We see a peak $|\pmb{v}_{\rm max}|/|\pmb{v}^{\rm post}_0|\approx2.3$ appears at $t\approx0.4t_{\rm shock}$. The amplification of pressure (i.e., the ratio of the maximum to the mean) reaches a slightly lower factor of $\approx1.5$. 

The difference of parallel and perpendicular shocks appears in magnetic field. For parallel shocks, the maximum magnetic field exponentially increases to $\approx30|\pmb{B}^{\rm post}_0|$ at $t\approx0.5t_{\rm shock}$ (see Fig.~\ref{fig:max_time}). The rapid increase of magnetic field comes from the stretching of magnetic fields by solenoidal turbulent motions induced in the postshock flow. The turbulent amplification of magnetic fields, i.e., turbulent dynamo, happens when magnetic fields are stretched by turbulent shear. In the kinematic stage, with both conservation of mass and magnetic flux, the turbulent stretching of a magnetic flux tube causes its lengthening and increase of the magnetic field strength  \citep{2005PhR...417....1B}. The dynamo growth of magnetic fields slows down in the next nonlinear stage and the field strength reaches maximum  $\approx65|\pmb{B}^{\rm post}_0|$. The magnetic field in the case of perpendicular shock is amplified more. Its maximum value gets $\approx210|\pmb{B}^{\rm post}_0|$, which is approximately three times larger than the parallel shock case. The higher amplification ratio in the perpendicular shock case is contributed by the compression of magnetic field at the shock front.

Fig.~\ref{fig:energy} shows the magnetic field and turbulent energies per unit mass as function of time. Our measurements use the maximum magnetic field strength and its corresponding density and turbulent velocity. To compare with the analytical theory in \cite{2016ApJ...833..215X}, we also calculate the magnetic energy growth by following their analytical expressions:
\begin{equation}
    \begin{aligned}
        \mathcal{E}_B&=\frac{B^2}{2\rho}\approx \mathcal{E}_{B_0}\exp{(2\Gamma_vt)}, ~\text{\rm linear stage},\\
        \mathcal{E}_B&=\frac{B^2}{2\rho}\approx \mathcal{E}_{\rm cr}+\frac{3}{38}\epsilon (t-t_{\rm cr}),~\text{\rm nonlinear stage},
    \end{aligned}
\end{equation}
where $\mathcal{E}_{B_0}=\frac{(B_0^{\rm post})^2}{2\rho_0^{\rm post}}$ is the linear stage's initial magnetic energy, $\mathcal{E}_{\rm cr}$ is the  magnetic energy at the beginning of the nonlinear stage, i.e., the magnetic energy at the end of the linear stage, $\epsilon=L_{\rm inj}^{-1}v_{\rm inj}^3$ is the turbulent energy transfer rate, $L_{\rm inj}$ and $v_{\rm inj}$ are the injection scale and injection velocity of turbulence, $t_{\rm cr}\approx0.4t_{\rm shock}$ is the beginning time of the nonlinear stage (or the end of the linear stage), and $\Gamma_v$ represents the largest eddy turnover rate \footnote{ $\Gamma_v=v_{\rm dis}/l_{\rm dis}=(l_{\rm dis}/L_{\rm inj})^{1/3}v_{\rm inj}/l_{\rm dis}$, where $l_{\rm dis}\approx L_x/50$ is the numerical dissipation scale of turbulence, $v_{\rm inj}\approx0.68v_{\rm shock}$, and $L_{\rm inj}\approx L_x/5$. See Sec.~\ref{ssec:turshe}.}. Linear stage and nonlinear stages here refer to the stages with negligible and important magnetic back-reaction, respectively. The nonlinear stage with significant magnetic backreaction on turbulence has a linear-in-time growth of magnetic energy. The growth rate is a small fraction of $\epsilon$ because of the reconnection diffusion of turbulent magnetic fields \citep{2016ApJ...833..215X}. Note that the magnetic field in the simulation has been re-scaled by a factor of $1/\sqrt{4\pi}$.  The magnetic energy in the parallel shock for both "Para." and "Kol." models shows a similar evolution trend as the theoretical prediction. At $t\approx 0.5 t_\text{shock}$, the turbulent dynamo undergoes a transition from the linear/kinematic stage with the exponential growth of magnetic energy to the nonlinear stage with the linear-in-time growth of magnetic energy. We note that the original theory derived in \citet{2016ApJ...833..215X} applies to a stationary continuously-driven Kolmogorov turbulence. Here we find that the shock-driven turbulence can have a more complicated spectral shape due to the multi-scale driving that evolves with time (see Fig.~ \ref{fig:bvspectrum}). This can cause some discrepancy between our numerical result and their analytical prediction. Moreover, as the shock-driven turbulence decays over the largest eddy turnover time (see \S~\ref{ssec:turshe}), there is not sufficient time for the nonlinear turbulent dynamo to reach full energy equipartition at the largest scale of turbulence, which requires approximately six largest eddy turnover time
\citep{2016ApJ...833..215X}.


The perpendicular shock case further involves the compression of the magnetic field and results in a higher magnetic energy value at an early time. Its exponential stage is more rapid than the parallel shock case. Importantly, we find due to the reconnection diffusion of turbulent magnetic fields \citep{2005AIPC..784...42L}, the magnetic energy under both compression and dynamo effects cannot exceed the turbulent energy of the largest eddy \citep{2016ApJ...833..215X,2020ApJ...899..115X}, as shown in Fig.~\ref{fig:bvspectrum} and Fig.~\ref{fig:energy}.


Moreover, Fig.~\ref{fig:mean_time} presents the ratio of mean magnetic field and its initial value as a function of $t$. The magnetic field is averaged over the postshock's turbulent shell (see Fig.~\ref{fig:vis.}). Unlike the maximum magnetic field, the mean field strength has a much milder growth with time and reaches $\approx2.0|\pmb{B}^{\rm post}_0|$ for a parallel shock (including both "Para." and "Kol." models) and $\approx7.9|\pmb{B}^{\rm post}_0|$ for a perpendicular shock. \cite{2007ApJ...663L..41G} reported a similar amplification factor using 2D simulations and particularly, \cite{2018PhRvL.120y1101F} reported in Cassiopeia A the magnetic field amplification reaches $\approx1.7|\pmb{B}^{\rm post}_0|$. This observation is consistent with our parallel shock cases. In addition, the corresponding histograms of amplified magnetic field strength in the postshock's turbulent shell is given in Fig.~\ref{fig:hist}. It is apparent that the strong magnetic field occupies only a small fraction in space with a low volume filling factor.  This is expected as the magnetic energy does not reach the full equipartition with the largest turbulent eddy, and accordingly the peak scale of magnetic energy spectrum is smaller than the largest turbulence scale (see Figs.~\ref{fig:bvspectrum} and \ref{fig:anisotropy_ratio}). In addition, we see the histogram of parallel shock cases (i.e., "Para." and "Kol.") becomes wilder at later time. The fraction of both weak ($<|\pmb{B}^{\rm post}_0|$) and strong magnetic field ($>|\pmb{B}^{\rm post}_0|$) increases as the magnetic field gradually gets amplified and turbulent. The change of histogram for perpendicular shock, however, is less significant. The weak magnetic field always have a relatively high fraction. This suggests that the strong magnetic field's intermittency is more significant in the presence of shock compression.

\subsection{Distributions of postshock turbulence $M_{\rm s}$ and $M_{\rm A}$}
The observational techniques to measure $M_{\rm A}$ \citep{Lazarian18,Hu19a,2021ApJ...910...88X} 
in supernova remnants 
can be used to test the dynamo theory and probe the turbulent amplification of magnetic fields near supernova shocks. Fig.~\ref{fig:mams} presents the distributions of averaged (along the $z$-axis) postshock magnetic field strength, $\langle M_{\rm s}\rangle$, and $\langle M_{\rm A}\rangle$ in the vicinity of shock front. The $\langle M_{\rm s}\rangle$ and $\langle M_{\rm A}\rangle$ of turbulence are defined as:
\begin{equation}
\begin{aligned}
    \langle M_{\rm s}\rangle&=\langle |\pmb{v}^{\rm post}-\pmb{v}^{\rm post}_0|/c_{\rm s}^{\rm post}\rangle,\\
    \langle M_{\rm A}\rangle&=\langle |\pmb{v}^{\rm post}-\pmb{v}^{\rm post}_0|\sqrt{\rho^{\rm post}}/|\pmb{B}^{\rm post}|\rangle,\\
    c_{\rm s}^{\rm post}&=\sqrt{\gamma \frac{P^{\rm post}}{\rho^{\rm post}}},
\end{aligned}
\end{equation}
where $c_{\rm s}^{\rm post}$ is the postshock sound speed and $\pmb{v}^{\rm post}$ represents the total postshock velocity of fluid. The initial uniform postshock velocity $\pmb{v}^{\rm post}_0$ is subtracted to get the turbulent velocity.

As shown in Fig.~\ref{fig:mams}, perpendicular shock can amplify the postshock magnetic field strength further due to the additional shock compression. The local maximum strength of the averaged magnetic field along the $z$-axis can reach $\approx15|\pmb{B}^{\rm post}_0|$, while the maximum of parallel shock's case achieves $\approx5|\pmb{B}^{\rm post}_0|$. Nevertheless, both cases have the same distribution of $\langle M_{\rm s}\rangle$ with a median value $\approx0.15$. This suggests that the postshock turbulence and sound speed are independent of magnetic fields. The two quantities are regulated only by the pre-shock density fluctuation's distribution. The distributions of $\langle M_{\rm A}\rangle$ appear differences. Due to a stronger magnetic field, the $\langle M_{\rm A}\rangle$ of the perpendicular shock case gets smaller with a median value $\approx7$.  $\langle M_{\rm A}\rangle$ in the parallel shock case (with a median value $\approx17$) increases when get close to the shock front. However, as shown in Fig.~\ref{fig:ma}, the perpendicular case shows an opposite trend that $\langle M_{\rm A}\rangle$ increases when the distance from the shock front increases. This lower value of $\langle M_{\rm A}\rangle$ around shock front is caused by the compression of magnetic field by shock wave. The compression amplifies the preshock magnetic field by a factor of $\approx4$ in a strong shock limit so that the corresponding $\langle M_{\rm A}\rangle$ is only $\approx1/4$ of the parallel shock's one. In the region away from the shock front ($x<550$, see Figs.~\ref{fig:ma} and \ref{fig:mams}), $\langle M_{\rm A}\rangle$ of both cases drops from 12 to 4. In this region, turbulence and magnetic field's fluctuation both decay (see Fig.~\ref{fig:vis.}). $\langle M_{\rm A}\rangle$ in this region is related to the fluctuations driven by reverse shocks. 

Moreover, in Fig.~\ref{fig:max_time}, we see that the maximum magnetic strength reaches $\approx65|\pmb{B}^{\rm post}_0|$ for parallel shock and $\approx210|\pmb{B}^{\rm post}_0|$ for perpendicular shock. However, Fig.~\ref{fig:mams}  shows that in the vicinity of the shock front, the $\langle|B|\rangle$ averaged along the $z-$axis ranges from $\approx0$ to $\approx15$ for perpendicular shock and from $\approx0$ to $\approx5$ for parallel shock. This significant difference suggests that the maximum magnetic strength has a low volume filling factor and thus the average is much smaller than the maximum. Accordingly, Fig.~\ref{fig:energy} shows that the maximum magnetic energy per unit mass can become comparable to the turbulent energy, suggesting local $M_{\rm A}\approx1$ for perpendicular shock. The $\langle M_{\rm A}\rangle$ averaged along the $z-$axis, however, ranges from $\approx7$ to $\approx15$ (see Fig.~\ref{fig:ma}). It means that in many positions along the $z-$axis, the magnetic energy is still much lower than turbulent energy. This gives an important implication for observational measurement, which can only give an averaged magnetic field strength. It implies that the magnetic field locally in 3D space can be much stronger.

\section{Discussion}
\label{sec:dis}
\subsubsection{Comparison with earlier 2D and 3D studies}
Earlier studies of magnetic field amplification frequently used 2D MHD simulations \citep{2007ApJ...663L..41G,2009ApJ...695..825I,2012ApJ...747...98G,2014MNRAS.439.3490M}. The 2D simulation, however, does not complete the picture. In turbulent dynamo, magnetic field lines are stretched to achieve amplification. In a 2D simulation, the magnetic field lines cannot be stretched along the $z$-axis, and therefore the amplification is relatively weak (see the Appendix.~\ref{appendix}). 

The studies of 3D magnetic field amplification were performed by \cite{2013ApJ...772L..20I} and \cite{2016MNRAS.463.3989J}. Their studies considered a Kolmogorov-type preshock density distribution. However, the observed density spectrum in media with supersonic turbulence is shallower than the Kolmogorov scaling \citep{1998A&A...336..697S,2000ApJ...543..227D,2006PhDT........19S,2009SSRv..143..357L,2012A&ARv..20...55H,2017ApJ...846L..28X,2018ApJ...856..136P,HLB20,2020ApJ...905..159X}. Therefore, we investigate the shallow density spectrum's effect on magnetic field amplification using 3D MHD simulations in this work. Compared with the shallow spectrum, the Kolmogorov spectrum of density fluctuations is characterised by large-scale density structures because the spectrum's ampltiude drops more quickly at large wavenumbers, i.e., relatively small-scale structures. These density contrasts result in slightly more compressive postshock turbulence. Moreover, probably because the preshock density spectrum is cut off at a small wavenumber $k=10$, given the limited numerical resolution, the difference between different density spectral shape is not obvious. The slopes of the postshock turbulence and magnetic field's spectrum in both cases (i.e., shallow and Kolmogorov preshock density spectra) are close to $-5/3$, i.e., the Kolmogorov scaling. One should also note that the density contrast at the same length scale for a Kolmogorov density spectrum in a warm medium and for a shallow density spectrum in a cold medium is very different, although here the same value is adopted in our simulations. In warm medium, the largest density fluctuation is even smaller than the mean, while in cold medium the density fluctuation can be much larger than the mean.

In addition, the injection scale of postshock turbulence is $k\lesssim10$ instead of $k=1$ in our simulations. This leads to a higher turbulent turnover rate and higher dynamo amplification because the fastest eddy turnover rate is responsible for magnetic energy's exponential growth in the initial kinematic stage. Nevertheless, due to the limited numerical resolution, the size of the smallest turbulent eddy in our simulations can be much larger than that in reality. Our linear stage of the dynamo is hence longer than that in reality, while the nonlinear stage is shorter \citep{2020MNRAS.496.5528M}.

\subsubsection{Effect of reconnection diffusion}
Turbulent dynamo is responsible for amplifying magnetic field, while reconnection diffusion has opposite effect. Reconnection diffusion causes the inefficient growth of magnetic energy during the nonlinear turbulent dynamo \citep{2016ApJ...833..215X}. Reconnection diffusion also limits the growth of magnetic energy up to the maximum turbulent energy, as shown in Fig.~\ref{fig:energy}.

Reconnection diffusion is also important for understanding star formation \cite{2005AIPC..784...42L}. Because of the reconnection diffusion, the magnetic field during the gravitational collapse of a cloud has a weaker dependence on density compared to the expectation under the flux-freezing assumption \citep{2020ApJ...899..115X}. The above prediction in \citep{2020ApJ...899..115X} has been numerically confirmed in e.g., \citet{2022MNRAS.511.5042S}.

\subsection{Implication on cosmic ray diffusion and acceleration}
The magnetic fields in postshock and preshock regions are critical in understanding the CRs diffusion and acceleration via the diffusive shock acceleration (\citealt{1977ICRC...11..132A,1978MNRAS.182..147B}). The amplified magnetic fluctuations in these regions act as magnetic scattering centers and magnetic mirrors that confine accelerated particles near the shock \citep{2022ApJ...925...48X}. The maximum attainable particle energy depends on the magnetic field strength \citep{1983A&A...125..249L}. This paper focuses on the postshock turbulence and magnetic field amplification. The preshock turbulence and turbulent magnetic fields can be generated in the shock precursor \citep{2009ApJ...707.1541B,2017ApJ...850..126X,2016MNRAS.458.1645D}. 

An accurate description of properties of magnetic field is required for describing CRs physics. For instance, the diffusion coefficient of CRs in directions parallel and perpendicular to magnetic fields both depends on properties of turbulent magnetic fields. Our results show that the perpendicular shock amplifies the magnetic field more significantly than parallel shock (see Fig.~\ref{fig:max_time}). In particular, the Alv\'en Mach number decrease in the vicinity of perpendicular shock's front (see Fig.~\ref{fig:ma}). A proper description on CR diffusion in different physical conditions is necessary for determining the CR acceleration efficiency \citep{2022ApJ...925...48X}.

Moreover, our results show the postshock turbulence and magnetic field are dominated by solenoidal component following the Kolmogorov scaling (see Fig.~\ref{fig:bvspectrum}). The scaling leads to that the CRs' perpendicular displacement increases with time to the power of 3/2 travelled along local magnetic field lines \citep{2013ApJ...779..140X,HLX21b}, when the mean free path is larger than turbulence's injection scale and scattering is inefficient. Such a superdiffusion of CRs in MHD turbulence arises from the accelerated separation of magnetic field lines. This behavior is predicted in \cite{LV99} and is supported by numerical simulations in \cite{2013ApJ...779..140X}. The many implications of this accelerated  separation of magnetic field lines include the modification of CR acceleration \citep{2014ApJ...784...38L}. In addition, we find the strong magnetic field has a low volume filling factor, which may help increase the scattering efficiency due to the large magnetic fluctuations.



\section{Summary}
\label{sec:con}
We investigate the interaction between shock wave and inhomogeneous density distribution in the preshock region by employing high-resolution three-dimensional MHD simulations. We consider two types of preshock density distribution. One is a shallow density spectrum with a 1D slope $\approx-0.5$, and the other is the Kolmogorov spectrum with a 1D slope $\approx-1.67$. The former is motivated by the observation of shallow density spectra in the Galactic disk. Due to the absence of preexisting preshock velocity fluctuation in the simulations, the postshock turbulence is driven only by the shock's interaction with the preshock inhomogeneous medium. The postshock turbulence further amplifies the postshock magnetic field via the turbulent dynamo. We summarize the main results as follows:
\begin{enumerate}
    \item We find that with a similar density contrast, a preshock shallow density spectrum is more efficient in generating the postshock turbulence than a Kolmogorov density spectrum.
    \item We find that solenoidal components dominate the postshock turbulence and postshock magnetic field. The postshock turbulence and magnetic field spectrum follow the Kolmogorov scaling. This is independent of the preshock density distribution's spectrum.
    \item We find that a preshock Kolmogorov density spectrum can lead to a slightly larger fraction of compressive turbulence compared to the case with a preshock shallow density spectrum.
    \item The postshock magnetic field can be by amplified up to a factor of $\gtrsim100$ in the presence of preshock shallow density spectrum. The nonlinear turbulent dynamo, therefore, can account for the magnetic field amplification by two orders of magnitude in young supernova remnants as indicated by observations.
    \item We provide numerical comparison of 2D and 3D simulations. We find that the magnetic field amplification is stronger and the magnetic field's morphology different.
    \item We test the nonlinear turbulent dynamo theory developed by \cite{2016ApJ...833..215X} and found a good agreement on the time evolution of magnetic energy during the nonlinear stage with their theoretical prediction. We show that the maximum magnetic energy is limited by the largest turbulent kinetic energy.
    \item We find that the strongest magnetic field achieved via turbulent dynamo has a low volume filling factor. The magnetic field strength averaged along the line-of-sight i.e., the $z$-axis in our simulations, is reduced by a factor of $\approx10$.
\end{enumerate}

\acknowledgments
S.X. acknowledges the support for this work provided by NASA through the NASA Hubble Fellowship grant \# HST-HF2-51473.001-A awarded by the Space Telescope Science Institute, which is operated by the Association of Universities for Research in Astronomy, Incorporated, under NASA contract NAS5-26555. A.L. and Y.H. acknowledges the support of NASA ATP AAH7546 and the allocation of computer time by the Center for High Throughput Computing (CHTC) at the University of Wisconsin-Madison. We acknowledge Ethan T. Vishniac for fruitful discussions.

\software{Athena++ code \citep{Stone20}}

\bibliographystyle{aasjournal}
\bibliography{Hu}

\appendix

\section{Comparing with 2D simulation}
\label{appendix}
\begin{figure}[htp]
    \centering
	\includegraphics[width=1.05\linewidth]{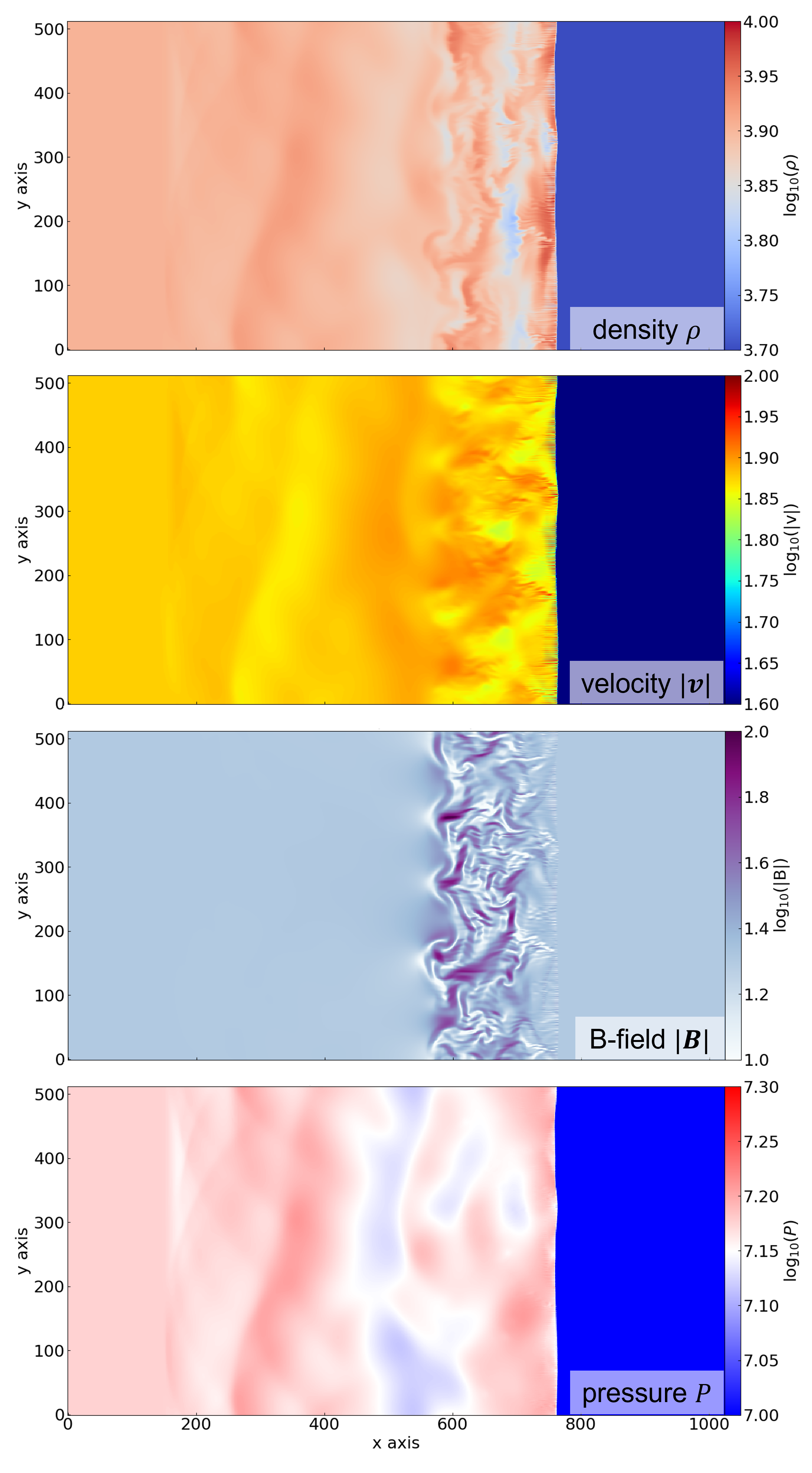}
    \caption{The maps of gas density, total velocity, magnetic field, and pressure. The 2D simulation cube $\beta=10$ at $t\approx0.75t_{\rm shock}$ is used here. Mean magnetic field is parallel to the $x$-axis. The physical variables are expressed in term of numerical units.}
    \label{fig:vis2d}
\end{figure}

\begin{figure}[h!]
    \centering
	\includegraphics[width=1.03\linewidth]{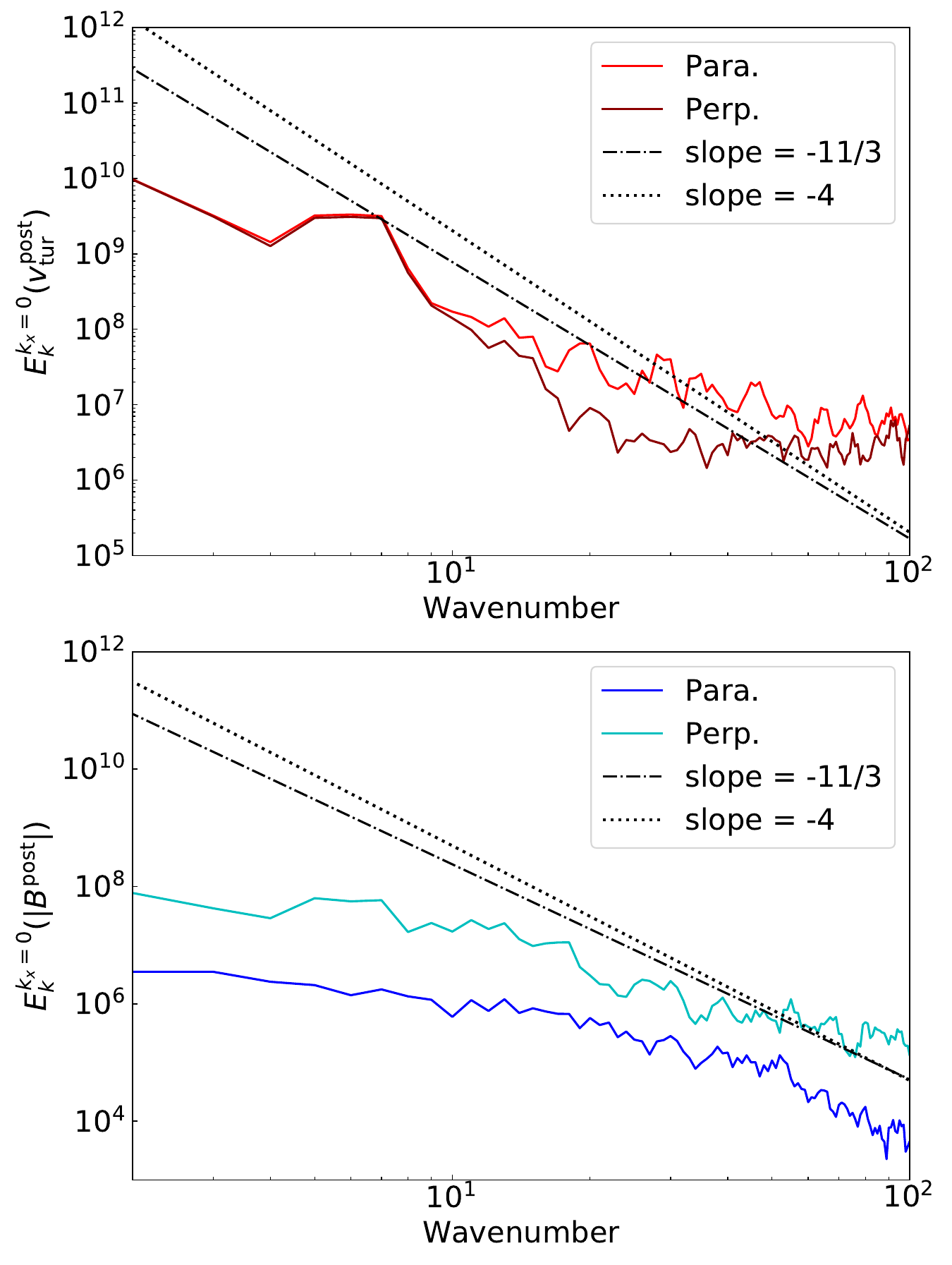}
    \caption{Comparison of energy spectrum of postshock turbulent magnetic field obtained from 3D simulation ($1024\times512\times512$ cells; bottom) and 2D simulation ($1024\times512\times1$ cells; top) at $t\approx0.75t_{\rm shock}$. Note we consider the spectrum only on the $k_x=0$ place in Fourier space. Therefore, we have the slope $\approx -4 + (D-1)$ and $\approx -11/3 + (D-1)$ for compressive and solenoidal turbulence, respectively. Here $D=2$ stands for 3D simulation and $D=1$ is for 2D simulation.}
    \label{fig:spectrum2D}
\end{figure}

We further run the simulation in 2D using the same setups, including the cases of parallel and perpendicular shock. The 2D simulations have resolution $1024\times512\times1$ cells. The 2D maps of gas density, total velocity, magnetic field, and pressure are shown in Fig.~\ref{fig:vis2d}. Compared with the 3D case (see Fig.~\ref{fig:vis.}), the magnetic field is amplified less in the 2D simulation. 

Fig.~\ref{fig:spectrum2D} present the postshock turbulent velocity nand magnetic field's energy spectrum obtained from 2D simulation ($1024\times512\times1$ cells) at $t\approx0.75t_{\rm shock}$. Also, to avoid the effect of non-periodic boundary condition along the $x$-axis, we only calculate the spectrum at $k_x=0$. The 2D simulations' spectrum has lower amplitude. The slopes, however, are still following the Kolmogorov scaling $E_k\approx-11/3$ for 2D simulation.

\end{document}